\documentclass[aps,twocolumn,showpacs,preprintnumbers,floatfix]{revtex4}
\usepackage{graphicx}
\usepackage{epsfig}
\usepackage{dcolumn}
\usepackage{amsmath}

\def\be{\begin{equation}}
\def\ee{\end{equation}}
\def\bea{\begin{eqnarray}}
\def\eea{\end{eqnarray}}
\def\bsp{\be\begin{split}}

\def\bes{\be  \begin{split}}

\def\epsl{\epsilon}

\newcommand{\Rmnum}[1]{\expandafter\@slowromancap\romannumeral #1@}

\begin{document}

\title{ Lattice study on $\eta_{c2}$ and X(3872)}
\author{\small
Yi-Bo Yang,$^{1}$ Ying Chen,$^{1,2}$\footnote{cheny@ihep.ac.cn} Long-Cheng Gui,$^{1,2}$, Chuan Liu,$^{3}$ \\
Yu-Bin Liu,$^{4}$ Zhaofeng Liu,$^{1,2}$ Jian-Ping Ma,$^{5}$ and Jian-Bo~Zhang~$^{6}$\\
(CLQCD Collaboration) } \affiliation{ \small $^1$~Institute of High Energy Physics, Chinese Academy
of Sciences, Beijing
100049, China \\
$^2$~Theoretical Center for Science Facilities, Chinese Academy of Sciences, Beijing 100049, China\\
$^3$~School of Physics and Center for High Energy Physics, Peking University, Beijing 100871, China\\
$^4$~School of Physics, Nankai University, Tianjin 300071, China\\
$^5$~Institute of Theoretical Physics, Chinese Academy of Sciences, Beijing 100080, China\\
$^6$~Department of Physics, Zhejiang University, Zhejiang 310027, China }

\begin{abstract}
Properties of $2^{-+}$ charmonium $\eta_{c2}$ are investigated in quenched lattice QCD. The mass of
$\eta_{c2}$ is determined to be $3.80(3)\,{\rm GeV}$, which is close to the mass of $D$-wave
charmonium $\psi(3770)$ and in agreement with quark model predictions. The transition width of
$\eta_{c2}\to \gamma J/\psi$ is also obtained with a value of $\Gamma=3.8(9)\,{\rm keV}$. Since the
possible $2^{-+}$ assignment to $X(3872)$ has not been ruled out by experiments, our results help
to clarify the nature of $X(3872)$.
\end{abstract}

\pacs{11.15.Ha, 12.38.Gc, 13.20.Gd, 14.40.Pq, 14.40.Rt } \maketitle

\section{Introduction}
Even though the charmoiumlike resonance $X(3872)$ has been established for several
years~\cite{x3872-1,Aub:2005, Aco:2004, Aba:2004} with $M_X=3871.68\pm 0.17\,{\rm MeV}$ and
$\Gamma_X<1.2\,{\rm MeV}$~\cite{Choi:2011, PDG:2012}, the very nature of it has not been fully
understood till now. Its even $C$ parity has been firmly established from its decay to $J/\psi
\rho$~\cite{Abu:2006} and to $J/\psi \gamma$~\cite{Aub:2006}. Further analysis of its decay angular
distribution also constrains its total quantum number $J^{PC}$ to be either $1^{++}$ or $2^{-+}$.
The discovery of $X(3872)$ has triggered quite a number of theoretical interpretations by assuming
a quantum number $1^{++}$, such as the radial excitation of $\chi_{c1}$, the $D\bar{D}^*$
molecule~\cite{Bignamini2009:molecule1, Voloshin2004:molecule2, Lee2009:molecule3}, a tetraquark
state~\cite{Maiani2005:tetra1, Dubnicka2010:tetra2, Dubnicka2011:tetra3}, etc.; however, none of
them can accommodate all the observed features of $X(3872)$. The situation became more complicated
when the {\it BABAR} Collaboration reported in 2009 that $2^{-+}$ is more favored by the study of
the decay angular distribution of the process $X(3872)\rightarrow J/\psi
\pi^+\pi^-\pi^0$~\cite{delAmoSanchez:2010jr}. In contrast, the same analysis by the Belle
Collaboration claims that both the $1^{++}$ and $2^{-+}$ assignments are consistent with their
data~\cite{Choi:2011, Lange2011:belle}. Another controversial result comes from the measurements of
the radiative decays of $X(3872)$. For the decay mode $X(3872)\to \gamma J/\psi$, the {\it BABAR}
and Belle collaborations reported consistent measurements~\cite{BaBar:2009, Belle:2011}:
 \begin{eqnarray}
 {\rm Br}(B^{\pm}&\rightarrow& X(3872)K^{\pm}){\rm Br}(X(3872)\rightarrow J/\psi \gamma)\nonumber\\
 &=&(2.8\pm 0.8 \pm 0.1)\times 10^{-6}~~~(BABAR),\nonumber\\
 {\rm Br}(B^{\pm}&\rightarrow& X(3872)K^{\pm}){\rm Br}(X(3872)\rightarrow J/\psi \gamma)\nonumber\\
 &=&(1.78^{+0.48}_{-0.44}\pm 0.12)\times 10^{-6}~~({\rm Belle}).
 \end{eqnarray}
With the world average value ${\rm Br}(B^{+}\rightarrow X(3872)K^{+})<3.2\times 10^{-4}$, one can
estimate the branch ratio ${\rm Br}(X(3872)\rightarrow J/\psi \gamma)> 0.9\%~(BaBar)$ or
$0.6\%~({\rm Belle})$. However, for the decay mode $X(3872)\to \gamma \psi'$, {\it BABAR} measured
a $3.4\pm 1.4$ times larger branch ratio~\cite{BaBar:2009}, but Belle found no
evidence~\cite{Belle:2011}. This large discrepancy should be reconciled by further experimental
measurements.

Theoretically, if we are constrained to its charmonium assignments, $X(3872)$ can be either the
radial excitation of $\chi_{c1}$ (if $1^{++}$), say, $\chi_{c1}'$, or the ${}^1D_2$ charmonium
$\eta_{c2}$ (if $2^{-+}$). The potential quark model predicts the mass of $\chi_{c1}'$ to be
$3925\,{\rm MeV}$~\cite{BGS2005:2Pmass}, which deviates from the mass of $X(3872)$ by about
$50\,{\rm MeV}$. There are also many lattice studies predicting a $\chi_{c1}'$ mass ranging from
$3850$ to $4060$ MeV~\cite{Okamoto:2002, Chen:2001, Chen:2007, Liu:2011}, but with various
uncertainties of their own, where the key difficulty is the challenging task of extracting the
excited states. As for the $\eta_{c2}$, the quark models usually predict the mass to be in the
range 3770 to 3830 MeV~\cite{BGS2005:2Pmass, BG2004:1Dmass, Badalian2000:1Dmass}, which is even
further away from the mass of $X(3872)$. This is also reinforced by recent lattice studies (and
this work). At any rate, the mass parameter should not be the unique criterion for the
interpretation of $X(3872)$; more information is definitely desired\--for example, the radiative
transition properties of $\chi_{c1}'$ and $\eta_{c2}$, which are theoretically accessible and
hopefully can shed some light on the nature of $X(3872)$.

In this work, we will focus on the study of the properties of $\eta_{c2}$, such as its mass and
radiative transition width to $J/\psi$. There are actually several phenomenological studies on this
topic~\cite{Jia:2010jn, Kalash2010, Ke:2011jf}, but they are rather model dependent. In contrast,
the lattice QCD approach, as a method from first principles, can provide information that is more
model independent. An additional technical advantage in the study of $\eta_{c2}$ on the lattice is
that it is the ground state in the $2^{-+}$ channel and is free from the uncertainty of the
extraction of excited states. In view of the notorious bad signal-to-noise ratio for $P$ and $D$
wave states, we adopt the quenched approximation so as to obtain large enough statistics for
precise physical quantities to be derived. As for the quenched approximation, even though long-term
experiences show that it is safe for charm quark systems, and the resultant uncertainties can be
small, we still take several steps to check this and be assured of our results. We first calculate
the spectrum of the ground state charmonia, such as $1S$, $1P$ states, and make sure that the
experimental spectrum patterns are reproduced. As for the radiative transitions, we choose the
transition mode of the tensor charmonium $\chi_{c2}$ to $J/\psi$ as a calibration of the systematic
uncertainties of our formalism by comparing our result to the experimental value. After that, we
continue to the study of the radiative transition of $\eta_{c2}$ to $J/\psi$. All the lattice
calculations are carried out on anisotropic lattices which are suitable to the study of heavy
particles. The numerical techniques are standard: the mass spectra are extracted from two-point
functions, and the multipole amplitudes contributing to the transition widths are derived from the
calculation of relevant three-point functions with a local electromagnetic current insertion. We
apply two anisotropic lattices with different lattice spacings to estimate the lattice artifacts
owing to the finite lattice spacing.

This work is organized as follows: The formalism for the calculation of radiative transition widths
on the lattice is briefly introduced in Sec.~\ref{section2}. In Sec.~\ref{section3} are the
numerical details, where the lattice setup, the extraction of mass spectrum and transition form
factors are explained, and the numerical results are presented. Section~\ref{section4} is the
conclusion and discussion. The theoretical derivation of the multipole form factors is described in
the Appendixes.

\section{Formalism}\label{section2}
As mentioned above, in this work we aim at the lattice calculation of the radiative transition rate
of $\eta_{c2}$ to $J/\psi$. The general radiative transition width of an initial particle $i$ to a
final particle $f$ is
\begin{eqnarray}\label{width}
\Gamma(i\rightarrow \gamma f)&=& \int d\Omega_q
\frac{1}{32\pi^2}\frac{|\vec{q}|}{M_{i}^2}\frac{1}{2J_i+1}\nonumber\\
&\times&  \sum\limits_{r_i,r_j,r_{\gamma}}\left|M_{r_i,r_j,r_{\gamma}}\right|^2,
\end{eqnarray}
where $\vec{q}=\vec{p}_i-\vec{p}_f$ is the decay momentum with the mass-on-shell value
$|q|=(M_i^2-M_f^2)/(2M_i)$, $M_{i}$ and $M_f$ are the masses of the particles $i$ and $f$, and
$M_{r_i,r_f,r_{\gamma}}$ is the transition amplitude with $r_i, r_f, r_\gamma$ being the
polarizations of $i$, $f$, and the photon, respectively. To the lowest order of QED, the amplitude
$M$ is expressed explicitly as
\begin{equation}
M_{r_i,r_f,r_\gamma}=\epsilon_{\mu}^*(\vec{q},r_\gamma)\langle f(\vec{p}_f,r_f)|j_{\rm
em}^{\mu}(0)|i(\vec{p}_i,r_i)\rangle,
\end{equation}
where $\epsilon_{\mu}^*(\vec{q},r_\gamma)$ is the polarization vector of the photon, and $\langle
f(\vec{p'},r_f)|j_{\rm em}^{\mu}(0)|i(\vec{p},r_i)\rangle$ gives the on-shell matrix elements of
the electromagnetic current $j_{\rm em}^\mu(x)=\bar{\psi}Q\gamma^\mu\psi(x)$ between the $i$ and
$f$ states. [Here $\psi$ refers to an array of all the contributing quark flavors, such as
$u,d,s,c,\ldots$, and $Q$ is a diagonal matrix of quark electric charges, say, ${\rm
diag}(Q)=Q_u,Q_d,Q_s,Q_c, \ldots$.] The hadronic matrix element can be derived directly from the
lattice QCD calculation of the related three-point functions:
 \begin{eqnarray}\label{eq_three}
\Gamma^{(3)\mu}_{mn}(\vec{p}_f,\vec{q},t,t')&&=\sum_{\vec{x},\vec{y}}
e^{-i\vec{p}_f\cdot\vec{x}}e^{-i\vec{q}\cdot\vec{y}}\times
\nonumber\\
&&\langle O_m^f(\vec{x},t)j_{\rm em}^\mu(\vec{y},t') O_n^{i\dagger}(\vec{0},0)\rangle,
\end{eqnarray}
where $O_{m,n}^{i,f}$ are the interpolating fields for the particles $i$ and $f$, with the indices
$m,n$ referring to different spatial components for spin nonzero states. The explicit derivation
can be expressed as
\begin{eqnarray}\label{three}
\Gamma^{(3),\mu}_{mn}(\vec{p}_f,\vec{q},t,t')&=&\sum_{r_i,r_f}e^{-E_f t}e^{-(E_i-E_f)t'}\nonumber\\
&\times&\frac{\hat{Z}_{m}^f(\vec{p}_f,r_f)\hat{Z}_{n}^{i*}(\vec{p}_i,r_i)}{2E_i\ 2E_f}\nonumber\\
&\times& \langle f(\vec{p}_f,r_f)|j_{\rm em}^\mu(0)|i(\vec{p}_i,r_i)\rangle\nonumber\\
&&(t',t-t'\rightarrow \infty),
\end{eqnarray}
where $\hat{Z}_{m}^{i,f}$ are the matrix elements like $\hat{Z}_{m}^X(\vec{p}_X,r_X)=\langle
0|O_m^{X}|X(\vec{p}_X,r_X)\rangle$, which can be derived from the relevant two-point functions,
\begin{eqnarray}\label{two}
\Gamma_{X,mn}^{(2)}(\vec{p}_X,t)&=&\sum\limits_{-\vec{x}}e^{i\vec{p}_X\cdot\vec{x}}\langle
O_{m}^X(\vec{x},t)O_{n}^{X\dagger}(\vec{0},0)\rangle\nonumber\\
&\rightarrow&\frac{1}{2E_X}e^{-E_X t}\sum\limits_{r_X}\langle
0|O_{m}^X|X(\vec{p}_X,r_X)\rangle\nonumber\\
 &\times&\langle X(\vec{p}_X,r_X)|O_{n}^{X\dagger}|0\rangle~~~(t\rightarrow \infty).
\end{eqnarray}
On the other hand, in the Minkowski space-time, the matrix elements  $\langle
f(\vec{p}_f,r_f)|j_{\rm em}^{\mu}(0)|i(\vec{p}_i,r_i)\rangle$ can be generally expressed by several
Lorentz-invariant form factors $F_k(Q^2)$ and Lorentz-covariant kinematic factors $\alpha_k(p_i,
p_f)$ through the multipole decomposition,
\begin{equation}\label{eq_ampth}
 \langle f(\vec{p}_f,r_f)|j_{\rm em}^{\mu}(0)|i(\vec{p}_i,r_i)\rangle=\sum_k \alpha_k^{\mu}(p_i,p_f)F_k(Q^2),
\end{equation}
where $p_{i,f}$ are now the four-momenta of particles $i$ and $f$, and $Q^2$ is the squared
transfer momentum $Q^2=-(p_i-p_f)^2$. Obviously, the concrete form factors $F_k(Q^2)$ and the
explicit expressions of the kinematic factors $\alpha_k$ depend on the properties of the particles
$i$ and $f$, and therefore should be worked out case by case. Finally, the decay width with an
on-shell photon ($Q^2=0$) can be expressed as
\begin{equation}
\Gamma(i\rightarrow \gamma f) \propto \sum\limits_k F_k^2(0).
\end{equation}
So the key problem in this work is to reliably extract these form factors through the lattice
calculation of the relevant hadronic two-point functions and three-point functions described above.

\section{Numerical Details}\label{section3}

We use the quenched approximation in this study. The gauge configurations are generated by the
tadpole improved gauge action~\cite{morningstar} on anisotropic lattices with the temporal lattice
much finer than the spatial lattice, say, $\xi=a_s/a_t\gg 1$, where $a_s$ and $a_t$ are the spatial
and temporal lattice spacings, respectively. The much finer lattice in the temporal direction
yields a higher resolution to hadron correlation functions, such that the masses of heavy particles
can be tackled on relatively coarse lattices. We have two anisotropic lattices ($L^3\times T=
8^3\times 96$ and $12^3\times 144$) with $\xi=5$. The relevant input parameters are listed in
Table~\ref{tab:lattice}, where the lattice spacings, say, $a_s=0.222(2)\,{\rm fm}$ for the coarser
lattice and $a_s=0.138(1)\,{\rm fm}$ for the finer lattice, are determined from
$r_0^{-1}=410(20)\,{\rm MeV}$ by calculating the static potential. For each lattice, we generate
1000 configurations, each of which is separated by 500 heat-bath updating sweeps to avoid the
autocorrelation. For fermions, we use the tadpole improved clover action for anisotropic
lattices~\cite{chuan1}. The parameters in the action are tuned carefully by requiring that the
physical dispersion relations of vector and pseudoscalar mesons are correctly reproduced at each
bare quark mass~\cite{chuan2}. The bare charm quark masses for the two lattices are set by the
physical mass of $J/\psi$ $m_{J/\psi}=3.097\,{\rm GeV}$.

In this work, we only consider the connected diagrams in the calculation of two-point and three-
point functions. The contribution of the disconnected diagrams is assumed to be small for
charmonium states due to the OZI suppression.

\begin{table}
\caption{\label{tab:lattice} Relevant input parameters for this work. The spatial lattice spacing
$a_s$ is determined from $r_0^{-1}=410(20)\,{\rm MeV}$ by calculating the static potential. }
\begin{ruledtabular}
\begin{tabular}{cccccc}
$\beta$ &  $\xi$  & $a_s$(fm) & $La_s$(fm)&
 $L^3\times T$ & $N_{conf}$ \\\hline
   2.4  & 5 & 0.222 & 1.78 &$8^3\times 96$ & 1000 \\
   2.8  & 5 & 0.138 & 1.66 &$12^3\times 144$ & 1000  \\
\end{tabular}
\end{ruledtabular}

\end{table}
\par
\subsection{Ground-state charmonium spectrum}
As the first step, we carry out a careful study on the ground-state charmonium spectrum, which can
illustrate to some extent the systematic uncertainties due to the quenched approximation. For the
states $\eta_c(0^{-+})$, $J/\psi(1^{--})$, $h_c(1^{+-})$, $\chi_{c0}(0^{++})$, and
$\chi_{c1}(1^{++})$, we adopt the conventional quark bilinear operators like $\bar{c}\Gamma c$,
with $\Gamma=\gamma_5, \gamma_i,\sigma_{ij}, 1$, and $\gamma_5\gamma_i$, respectively. For the
tensor mesons $\chi_{c2}(2^{++})$ and $\eta_{c2}(2^{-+})$, since there are not quark bilinear
operators, we build the corresponding operators by combining the quark bilinear operator with
either the spatial gauge-covariant derivatives $D_i$ or the color magnetic field strength operator
$B_i$, which is built from Wilson loops. It is known that the spin $J=2$ states in the continuum
correspond to both the $T_2$ and $E$ irreducible representations (irreps) of the cubic point group
 $O$ on finite lattices, so the interpolating field operators of the two irreps are constructed for
 the tensor charmonia. For example, the $T_2$ operator for the $\chi_{c2}(2^{++})$ state is taken as
$|\epsilon_{ijk}|\bar{c}\gamma_j \overleftrightarrow{D}_k c$ where
$\overleftrightarrow{D}=\overleftarrow{D}-\overrightarrow{D}$, and the $E$ operator is also built
to check the restoration of the continuum rotation symmetry.

We will emphasize the choice of the operators for the $2^{-+}$ state, which is the major object of
this work.  The situation for the $\eta_{c2}$ meson is a little bit more complicated. We try first
three types of operators, such as
\begin{eqnarray}\label{eq_opera}
&&|\epsilon_{ijk}|\bar{c}(x)\Sigma_j\overleftrightarrow{D}_k c(x)~~(D{\rm -type}),\nonumber\\
&&|\epsilon_{ijk}|\bar{c}(x)\gamma_5 \overleftrightarrow{D}_j \overleftrightarrow{D}_k c(x)~~(DD{\rm -type}),\nonumber\\
&&|\epsilon_{ijk}|\bar{c}(x)\gamma_j B_k c(x)~~(F{\rm -type}),\nonumber
\end{eqnarray}
where only the $T_2$ operators are presented ($E$ operators can be built similarly, and the details
can be found in Ref.~\cite{Liao:2002rj}).

It is known that the signal-to-noise ratios of the correlation functions are always bad for
$P$-wave and $D$-wave states. To circumvent this difficulty, we adopt the Coulomb-gauge fixed wall
source techniques in the calculation of the spectrum. The configurations are fixed to the Coulomb
gauge first, then the charm quark propagators are calculated with uniform wall source vectors. For
the spin $J=0,1$ states, the point-sink wall source correlation functions can be constructed
straightforwardly with these propagators. For the tensors, we use the $F$-type operators as the
wall source, which means that additional inversions should be carried out with wall sources
multiplied by the local color field strength operators $B_i(x)$. On the other hand, since the gauge
is fixed, the gauge-covariant derivative operator $\overleftrightarrow{D}$ is replaced by the
direct derivative operator
$\overleftrightarrow{\nabla}=\overleftarrow{\nabla}-\overrightarrow{\nabla}$ in the practical
calculation.

\begin{figure}
\includegraphics[scale=0.6]{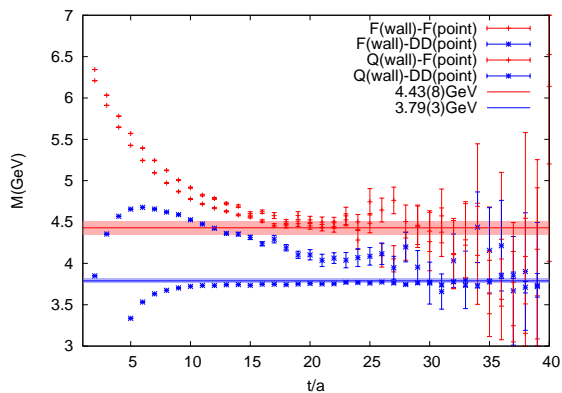}
\caption{\label{fig_2mp_mass} The $2^{-+}$ effective masses of $F-F$, $DD-F$, $F-Q$ and $DD-Q$
correlators at $\beta=2.8$ are plotted for illustration. $F-F$ and $F-Q$ effective masses lie on
each other and saturate to the same plateau with the best-fit mass $M=4.43(8)\,{\rm GeV}$
(indicated by the red line with the jackknife error band). The $DD-Q$ effective mass has a plateau
at $M=3.79(3)$ ( blue line with the jackknife error band ). In contrast, $DD-F$ effective mass does
not show perfect plateau, but it evolves gradually from the upper plateau to the lower. }
\end{figure}

The masses of $1S$ and $1P$ charmonium states can be neatly derived with the standard data
analysis, however, the situation for the $2^{-+}$ channel is very strange.
Figure~\ref{fig_2mp_mass} shows the effective masses of various correlation functions of this
channel at $\beta=2.8$. It is seen that the effective mass of the $F$-type point sink and $F$-type
wall source correlator ($F-F$) saturate at a plateau with the best-fit mass $4.43(8)\,{\rm GeV}$,
while that of the $DD$-type point sink and $F$-type wall source correlator ($DD-F$) goes lower and
does not show a perfect plateau. Intuitively, a mass of $4.4\,{\rm GeV}$ is too large for the
$2^{-+}$ ground state charnonium. Thus what one can infer from these behaviors is that the $F$-type
operator couples predominantly to a higher state but little to the conventional charmonium; in the
mean time, there must be a lower state which can be accessed by the $DD$-type operator but whose
spectral weight is relatively small due to the $F$-type wall source. To check this and to dig out
the desired $2^{-+}$ charmonium state, we try instead another wall source operator ($T_2$ irreps
for example),
$$|\epsilon_{ijk}|\sum\limits_{\vec{x},\vec{y},\vec{z}}\bar{c}^a(\vec{x},0)\gamma_j
c'^a(\vec{x},0)\bar{c'}^b(\vec{y},0)\Sigma_k c^b(\vec{z},0)~~(Q-{\rm type}),$$ where
$\Sigma_k=\epsilon_{ijk}\sigma_{ij}$ and $c'$ stands for a quark field with the same mass as that
of the charm quark but a different flavor. For simplicity, we call this operator $Q$-type in the
context. With this type of wall source operator, the effective masses of the $F$-type point sink
correlator ($F-Q$) and the $DD$ type point sink correlator ($DD-Q$) are also plotted in
Fig.~\ref{fig_2mp_mass}, where one can find that the mass plateau of the $F-Q$ correlator coincides
with that of the $F-F$ correlator within errors, while the effective mass of the $DD-Q$ correlator
shows a very nice plateau with the best-fit mass $3.79(3)\,{\rm GeV}$. Since the lower state has a
mass close to the potential model prediction of $2^{-+}$ charmonium and the higher state is much
heavier, we assign the lower state to the conventional ${}^1D_2$ charmonium state $\eta_{c2}$. This
assignment can be reinforced by comparison with the established $1{}^3D_1$ charmonium state
$\psi(3770)$: They are both $D$-wave charmonia and are therefore close in mass; the small mass
splitting can be attributed to the different spin-spin and spin-orbital interactions.

\begin{figure}
\includegraphics[scale=0.6]{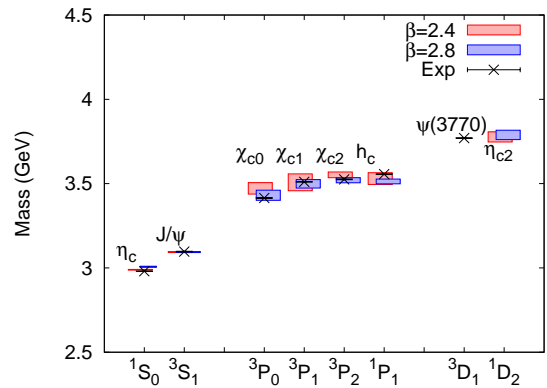}
\caption{\label{spectrum} $1S$, $1P$, and $1D$ charmonium spectrum. The red boxes illustrate the
results for $\beta=2.4$, and the blue ones for $\beta=2.8$. The experimental value are also plotted
with points for comparison.}
\end{figure}

The whole spectrum of the lowest-lying charmonium states we extracted in this work is illustrated
in Fig.~\ref{spectrum} and listed in Table~\ref{mass_table}, where the experimental values are also
given for comparison. Since we have only two lattice spacings, we would not carry out a serious
extrapolation to the continuum limit, but we show all the results, from which one can see that the
effects of the finite lattice artifacts and the quenched approximation are not that important.
\begin{table}
\caption{\label{mass_table} Listed here are the masses of the lowest-lying charmoinum states
extracted from the two lattices ($\beta=2.4$ and $\beta=2.8$) in this work. The experimental
results~\cite{PDG:2012} and the nonrelativistic quark model predictions~\cite{BGS2005:2Pmass} are
also given for comparison.}
\begin{ruledtabular}
\begin{tabular}{cccccc}
meson              & $J^{PC}$ & $M (2.4)$  & $M(2.8)$ &  Expt. & QM

\\\hline
$\eta_c({}^1S_0)$    & $0^{-+}$ &  2.989(2)        &  3.007(3)       & 2.981  & 2.982\\
$J/\psi({}^3S_1)$    & $1^{--}$ &  3.094(3)        &  3.094(3)       & 3.097  & 3.090 \\
&&&&&\\
$h_c({}^1P_0)      $ & $1^{+-}$ &  3.530(35)       &  3.513(14)      & 3.526  & 3.516 \\
$\chi_{c0}({}^3P_0)$ & $0^{++}$ &  3.472(34)       &  3.431(30)      & 3.415  & 3.424 \\
$\chi_{c1}({}^3P_1)$ & $1^{++}$ &  3.508(50)       &  3.499(25)      & 3.511  & 3.505 \\
$\chi_{c2}({}^3P_2)$ & $2^{++}$ &  3.552(17)       &  3.520(15)      & 3.556  & 3.556 \\
&&&&&\\
$\psi''({}^3D_1)   $ & $1^{--}$ &     \--          &   \--           & 3.770  & 3.785 \\
$\eta_{c2}({}^1D_2)$ & $2^{-+}$ &  3.777(30)       &  3.789(28)      &   \--  & 3.799 \\
\end{tabular}
\end{ruledtabular}
\end{table}

The goal of the spectroscopy study in this work is twofold. First, the physical spectrum of
experimentally established charmonium states can be well reproduced in our formalism. This gives us
confidence in our prediciton of the $\eta_{c2}$ mass. Second, the practical study finds that the
$DD$-type operator is preferable for producing the $2^{-+}$ charmonium. Therefore, in the study of
its radiative transition, we choose the $DD$-type operator for $\eta_{c2}$ in the calculation of
the related three-point functions.

\par

\subsection{Renormalization of the vector current}

\begin{figure}[tbh]\label{fig_renorm}
\includegraphics[scale=0.6]{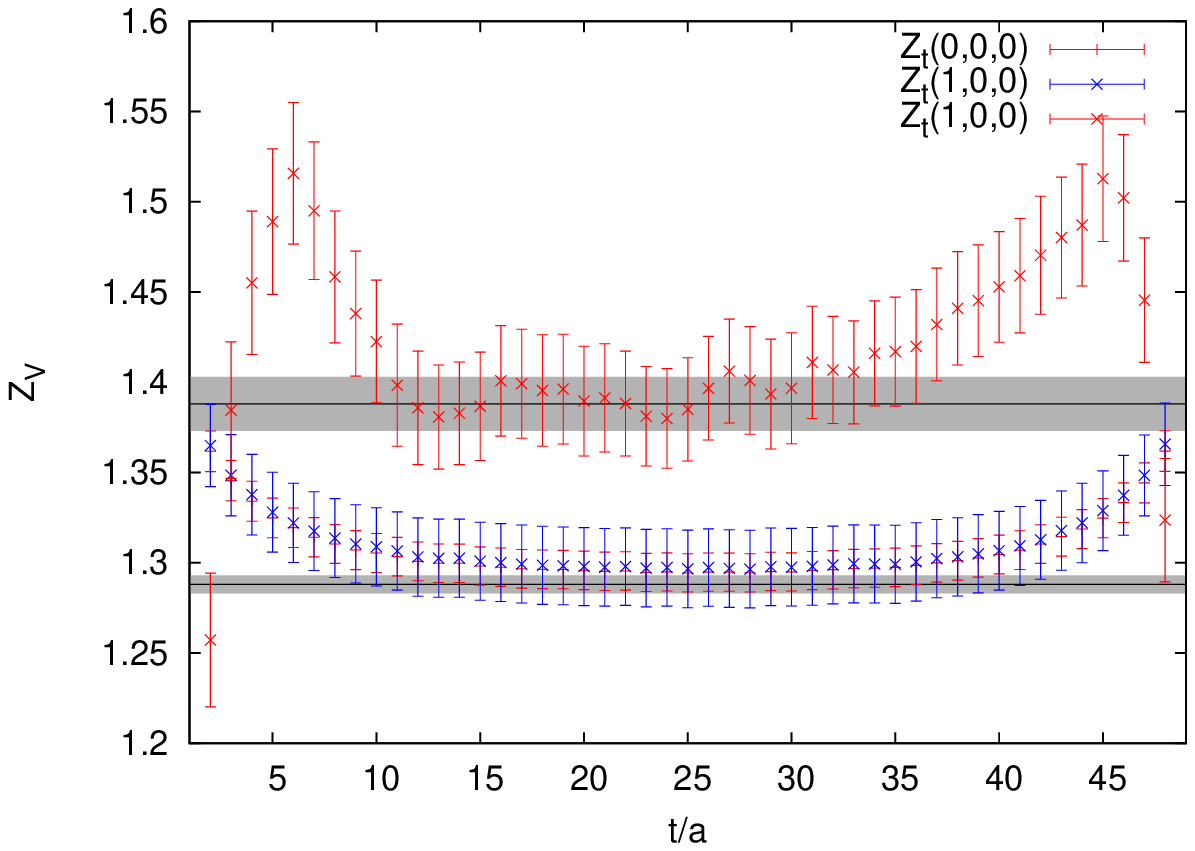}
\includegraphics[scale=0.6]{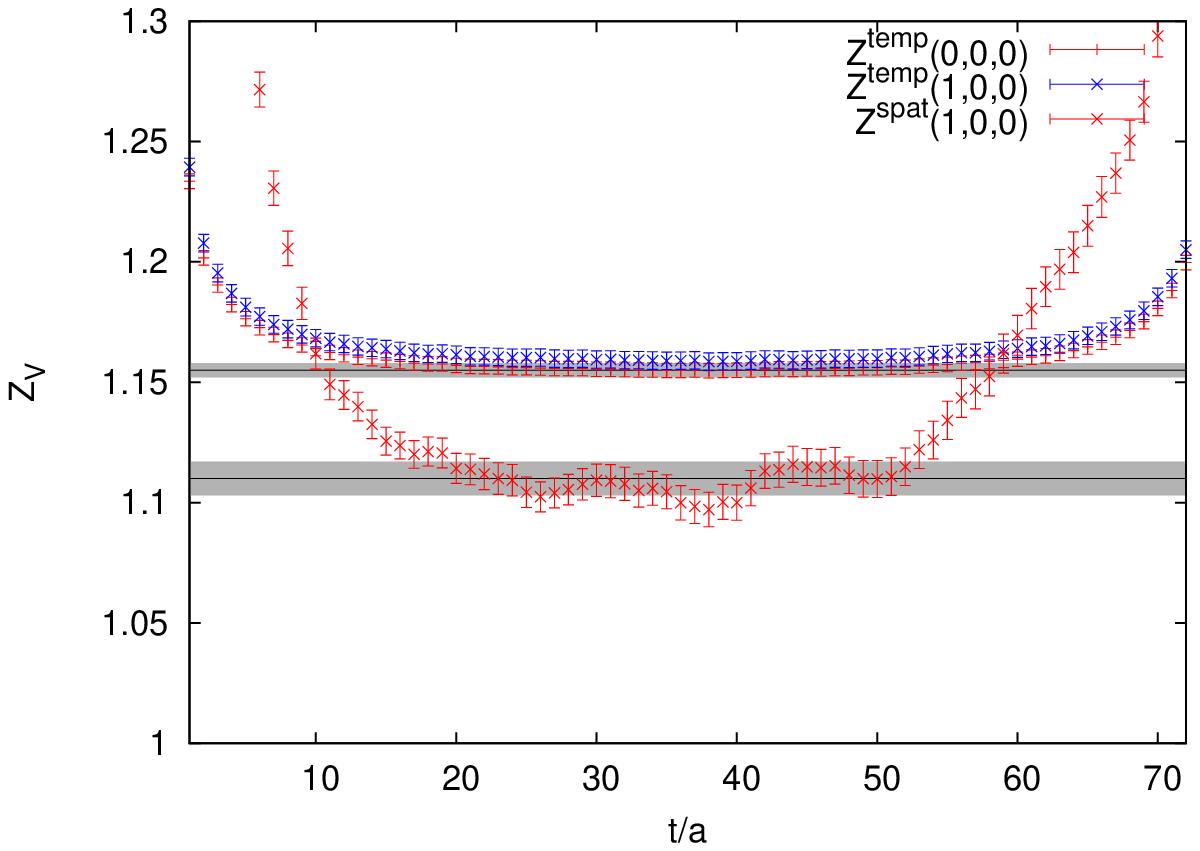}
\caption{\small The renormalization constant $Z_V$ of the vector current. The upper panel is for
$\beta=2.4$ and the lower one for $\beta=2.8$. The dots are the simulation data with jackknife
errors, and the lines show the fit results.}
\end{figure}

In the quenched approximation, since there are no sea quarks, the electromagnetic current
contributing to the radiative transitions of charmonia involves only the charm quark, say, $j_{\rm
em}(x)=Q_cj^\mu(x)$ with $j^\mu(x)=\bar{c}\gamma_\mu c (x)$, which is the one we adopt in this
study. It is a conserved vector current and need not be renormalized in the continuum. However, on
a finite lattice, it is not conserved anymore due to the lattice artifact and receives a
multiplicative renormalization factor $Z_V(a_s)$. Following the scheme proposed by
Ref.~\cite{Dudek:2006ej}, $Z_V(a_s)$ is extracted using the ratio of the $\eta_c$ two-point
function and the related three-point function evaluated at $Q^2=0$,
\begin{equation}
  Z^{(\mu)}_V(t) = \frac{p^\mu}{E(\vec{p})} \frac{\frac{1}{2} \Gamma^{(2)}_{\eta_c}(\vec{p};
  t_f=\frac{n_t}{2}) }
  {\Gamma^{(3),\mu}(
    \vec{p},\vec{p},\frac{n_t}{2}, t)}, \nonumber
\end{equation}
where the factor $1/2$ accounts for the effect of the temporal periodic boundary condition, and the
superscript $\mu$ of $Z_V(a_s)$ is used to differentiate the temporal component from the spatial
ones, since they are not necessarily the same due to the anisotropic lattices we use.
Figure~\ref{fig_renorm} plots $Z^{(\mu)}_V(t)$ with respect to $t$ for the two lattices.  $Z_V$'s
are extracted from the plateaus and the values are listed in Table~\ref{renorm}. Obviously, the
spatial components $Z_V^{(s)}(a)$ deviate from the temporal ones by a few percent. This deviation
can be attributed to the imperfect tuning of the bare velocity in the fermion action. In this work,
only $Z_V^{(s)}$'s enter the calculation since only the spatial components of the vector current
are involved in the extraction of the form factors.
\begin{table}
\caption{The renormalization constants $Z_V^{(s)}$ and $Z_V^{(t)}$ of the spatial and temporal
components of the vector current for $\beta=2.4$ and $\beta=2.8$ lattices. Two momentum modes,
(0,0,0) and $(1,0,0)$, are used for the derivation.}\label{renorm}
\begin{ruledtabular}
\begin{tabular}{cccc}
$\beta$ & $Z_V^{(t)}(0,0,0)$ & $Z_V^{(t)}(1,0,0)$ & $Z_V^{(s)}(1,0,0)$  \\\hline
 $2.4$    & $1.288(5)$  & $1.299(11)$   & $1.388(15)$ \\
 $2.8$    & $1.155(3)$  & $1.159(3) $   & $1.110(7)$
\end{tabular}
\end{ruledtabular}
\end{table}

\subsection{Three-point functions and form factors}
With the prescriptions discussed above, we now give a brief description of the calculation of the
three-point functions. In practice, we use local sink and source operators for the initial and the
final states, and insert the vector current $j^\mu(x)=\bar{c}\gamma^\mu c (x)$ only on the quark
line. (The current insertion on the antiquark line is numerically equivalent and is taken into
consideration by multiplying by a factor of 2 in the final result.) The three-point functions
contributed by the connected diagrams (disconnected diagrams are neglected) are calculated by using
the standard sequential source technique. (One can refer to Refs.\cite{Dudek:2006ej,bonnet05} for
the details.) In order to increase the statistics, we repeat the same calculations $T$ times (where
$T$ is the temporal lattice size) by setting a point source on a different time slice each time.
With the related two-point functions calculated accordingly, a straightforward way to extract the
interested matrix elements $\langle f(\vec{p'},r_f)|j^{\mu}(0)|i(\vec{p},r_i)\rangle$ is to fit the
three-point function and two-point function simultaneously according to Eq.~(\ref{three}) and
Eq.~(\ref{two}). However, it is known that the excited states contribute much to two-point and
three-point functions when the time ranges $t$ and $t_f-t$ are not large enough. This situation is
more serious for local operators, so it is not trivial to isolate the contribution of ground
states. A way around this is to employ the ratios of correlation functions, which can suppress the
contribution of excited states substantially. For this purpose, we introduce the functions
$R^\mu(t)$,
\begin{eqnarray}\label{eq_amp}
 &&R^\mu(t)=\Gamma^{(3)}(\vec{p_f},\vec{q},t_f,t)\times \nonumber\\
 &&\sqrt{\frac{2E_i\Gamma^{(2)}_i(\vec{p_i},t_f-t)}
    {\Gamma^{(2)}_i(\vec{p_i},t)\Gamma^{(2)}_i(\vec{p_i},t_f)}}
    \sqrt{\frac{2E_f\Gamma^{(2)}_f(\vec{p_f},t)}
    {\Gamma^{(2)}_f(\vec{p_f},t_f-t)\Gamma^{(2)}_f(\vec{p_f},t_f)}},\nonumber\\
\end{eqnarray}
which should be insensitive to the variation of $t$ in a time window, so that the desired matrix
element $ \langle f(\vec{p}_f,r_f)|j^{\mu}(0)|\psi|i(\vec{p}_i,r_i)\rangle$ can be extracted from
the plateau.

In the data analysis, we divide the 1000 configurations into 100 bins and use each bin average as
an independent measurement. For the resultant 100 bins, we use the one-eliminating jackknife
method. Since the energies $E_{i,f}$ can be determined very precisely from the two-point functions,
they are treated as known parameters in the above equation. Practically, $R^\mu(t)$ is fitted by
the function
\begin{equation}
R^\mu (t) = \langle f(\vec{p}_f,r_f)|j^{\mu}(0)|\psi|i(\vec{p}_i,r_i)\rangle+\delta f(t)
\end{equation}
where the additional term $\delta f(t)= a e^{-\delta m t}$ accounts for the residual contribution
of excited states.
Thus we can obtain a jackknife ensemble of the matrix elements. The second step of data analysis is
to extract the form factors that enter the calculation of decay widths. Since these matrix elements
can be expressed in terms of form factors through the multipole decomposition,
\begin{equation}
 \langle f(\vec{p}_f,r_f)|j^{\mu}(0)|i(\vec{p}_i,r_i)\rangle=\sum_k \alpha_k^\mu(p_i,p_f)\hat{F}_k(Q^2),
\end{equation}
and $\alpha^\mu_k (p_i,p_f)$ are theoretically known kinematic functions, the form factors
$\hat{F_k}(Q^2)$ can then be derived straightforwardly. Taking into consideration the contribution
of the current insertion on the antiquark line, the electric charge of the charm quark $Q_c=2/3$,
and the renormalization constant of the spatial components of the current operator $Z_V^{(s)}$,
$\hat{F_k}$ is related to $F_k$ of Eq.~(\ref{eq_ampth}) as
\begin{equation}
F_k(Q^2) = 2\times \frac{2}{3}e \times Z_V^{(s)} \hat{F}_k(Q^2).
\end{equation}
With this in mind, in the following context, we omit the hat of $\hat{F}$ and insert $Z_V^{(s)}$
implicitly in possible expressions.

In order to take good care of the correlation between the form factors, we carry out correlated
minimal $\chi^2$ fits with the jackknife covariance matrix built from the jackknife ensemble of the
matrix elements. On the other hand, for a specific $Q^2$, there may be several symmetric copies of
the matrix elements with the same value of $\alpha_k^\mu$. These copies are averaged over to
increase statistics.

In the following subsections, we present first the calculation of the process $\chi_{c2}\rightarrow
\gamma J/\psi$ to see how precisely the form factors\-- and thereby the transition width\-- can be
derived, and then the results of $\eta_{c2}\rightarrow \gamma\psi$.

\subsection{$\chi_{c2}\rightarrow \gamma J/\psi$ transition}
The Minkowski space-time matrix elements for this transition can be expressed in terms of form
factors as follows:
\begin{eqnarray}
&&\langle V(\vec{p}_V, \lambda_V) | j^{\mu}(0) | T(\vec{p}_T, \lambda_T)\rangle = \alpha_1^\mu
E_1(Q^2) \nonumber\\
&&+ \alpha_2^{\mu}M_2(Q^2) + \alpha_3^\mu E_3(Q^2) + \alpha_4^\mu C_1(Q^2)+ \alpha_5^\mu
C_2(Q^2)\nonumber\\
\end{eqnarray}
where $V$ stands for the $1^{--}$ vector meson $J/\psi$, $T$ stands for the $2^{++}$ tensor
$\chi_{c2}$, and $\alpha_i^\mu$ are Lorentz covariant kinematic functions of $p_V$ and $p_T$ (and
specific polarizations of $V$ and $T$), whose explicit expressions are tedious and omitted here.
Although a $J=2$ representation of the rotational symmetry in the continuum breaks into the $E$ and
$T_2$ irreducible representations (irreps) of the lattice spatial symmetry group $O$, we find that
this breaking effect is small in our work, as is manifested by the near degeneracy of the masses
and spectral weights of the ground states in these two irreps when we study the relevant two-point
functions. Thus, we assume that the rotation symmetry breaking is also negligible for the related
matrix elements, and we carry out the multipole decomposition on the basis of $E\bigoplus T_2$,
which is equivalent to the $J=2$ basis up to an orthogonal transformation. One can find the
detailed decomposition procedure in Appendix~\ref{sec_a_pole2pp} and may also refer to
Refs.~\cite{Dudek:2006ej, Dudek2009}. In the practical study, we set $T$ to be at rest and let $V$
move with different spatial momenta $\vec{p}=2\pi \vec{n}/L$. The 27 momentum modes of
$\vec{n}=(n_1,n_2,n_3)$ ranging from $(0,0,0)$ to $(2,2,2)$ are calculated for $V$.

\begin{figure}
\includegraphics[scale=0.6]{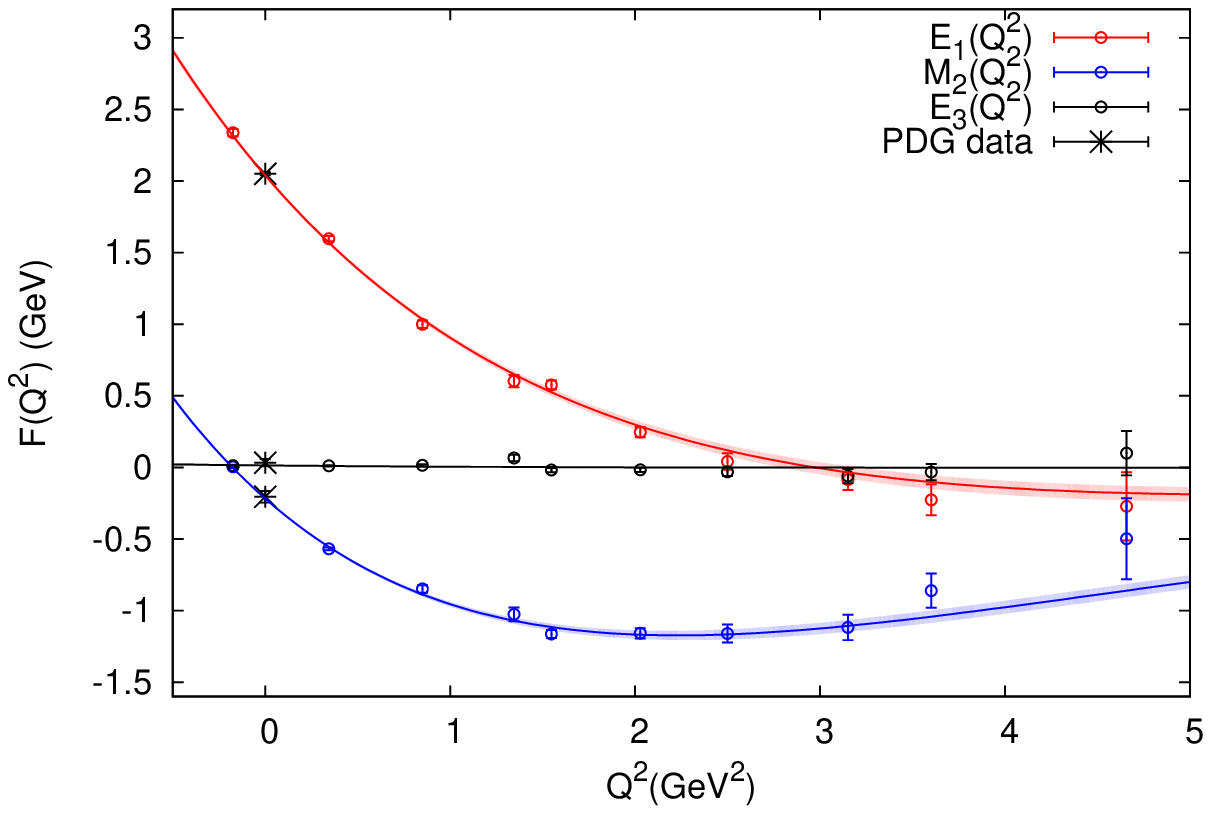}
\includegraphics[scale=0.6]{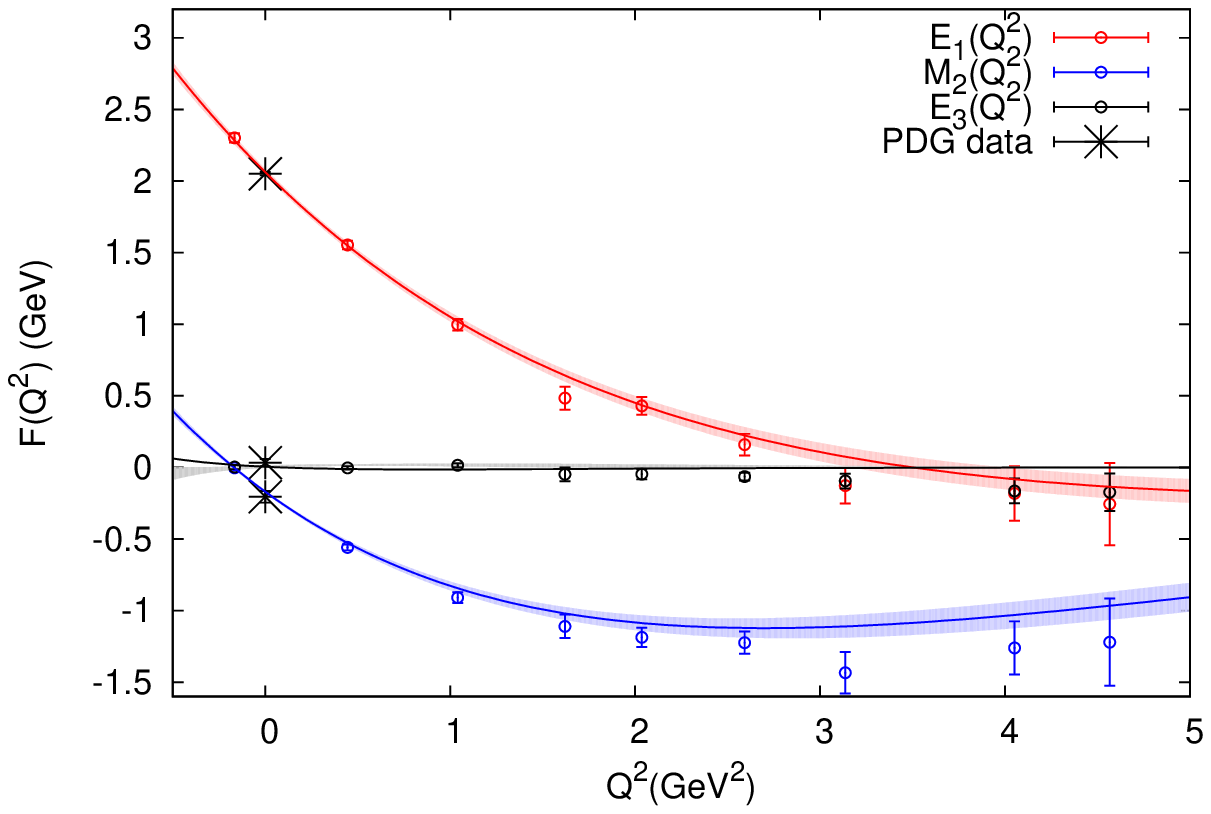}
\caption{\label{fig_2pp} The extracted form factors $E_1(Q^2)$, $M_2(Q^2)$, and $E_3(Q^2)$ are
plotted versus $Q^2$ for the two lattices of $\beta=2.4$ (the upper panel) and $\beta=2.8$ (the
lower panel), respectively, where the points are the simulation data, the line the fit function,
and the error bands the jackknife ones. The PDG values of $E_1(0)$ and $M_2(0)$ are also plotted
for comparison.}
\end{figure}

The transition width of $\chi_{c2}\rightarrow \gamma J/\psi$ for an on-shell photon ($Q^2=0$)
involves only the form factors $E_1(0)$, $M_2(0)$ and $E_3(0)$, or explicitly,
\begin{equation}
\Gamma(\chi_{c2}\rightarrow \gamma
J/\psi)=\frac{16\alpha|\vec{k}|}{45M^2_{\chi_{c2}}}(|E_1(0)|^2+|M_2(0)|^2+|E_3(0)|^2)\label{eq_form}
\end{equation}
where $|\vec{q}|=(M_{\chi_{c2}}^2-M_{J/\psi}^2)/2M_{\chi_{c2}}$ is the decaying energy of the
photon, and $\alpha=e^2/(4\pi)$ is the fine structure constant. Since our simulation data are
obtained at $Q^2\neq 0$, these on-shell form factors should be interpolated to $Q^2=0$. In doing
this, we adopt the fitting functional form inspired by the nonrelativistic quark
model~\cite{Dudek:2006ej},
\begin{equation}
F_k(Q^2)=F_k(0)(1+\lambda_k Q^2)e^{-\frac{Q^2}{16 \beta_k^2}}
\end{equation}
which has been applied successfully in previous works. Here $F_k(0)$, $\lambda_k$, and $\beta_k$
are the parameters to be fitted though a correlated $\chi^2$ fitting procedure where the covariance
matrix is constructed using the one-eliminating jackknife method. Plotted in Fig.~\ref{fig_2pp} are
the extracted form factors $E_1(Q^2)$, $M_2(Q^2)$, and $E_3(Q^2)$ versus $Q^2$ for the two lattices
of $\beta=2.4$ (the upper panel) and $\beta=2.8$ (the lower panel). The data points are the
simulation results, and the lines are the fit function with the jackknife error bands. One can find
that the data are very precise owing to the high statistics, and the fit errors are also very
small. We also carry out a simple polynomial fit with respect to $Q^2$, $F_k(Q^2)=F_k(0)+c_1 Q^2 +
c_2 Q^4$, and get consistent results within errors. Table~\ref{tab_2pp} lists the results of the
interpolation, where the continuum limit extrapolation is also given. It is seen that the electric
dipole ($E_1$) contribution is dominant in the transition $\chi_{c2}\rightarrow \gamma J/\psi$,
while the contribution of the magnetic quadrupole ($M_2$) is drastically suppressed, as depicted by
the ratio
\begin{equation}
a_2=\frac{M_2(0)}{\sqrt{E_1(0)^2+M_2(0)^2+E_3(0)^2}},
\end{equation}
for which we get a result $a_2=-0.107(3)$ for $\beta=2.4$ and $a_2=-0.082(7)$ for $\beta=2.8$.
After a linear extrapolation in $a_s^2$, we get the value in the continuum limit $a_2=-0.067(7)$,
which is consistent with the PDG data, where $a_2=-0.100\pm 0.015$~\cite{PDG:2012}. The
contribution of the electric octupole $E_3$ is far smaller. For the ratio
\begin{equation}
a_3=\frac{E_3(0)}{\sqrt{E_1(0)^2+M_2(0)^2+E_3(0)^2}},
\end{equation}
we obtain $a_3=0.007(2)$ for $\beta=2.4$, $a_3=0.003(4)$ for $\beta=2.8$, and the continuum limit
$a_3=-0.003(6)$, which are also compatible with the PDG data $a_3=0.016\pm 0.013$~\cite{PDG:2012}.
If we focus on $E_1$, we get the fitting parameters $\beta_1$ and $\lambda_1$,
\begin{eqnarray}
\beta_1&=&0.431(5)\,{\rm GeV}\nonumber\\
\lambda_1&=& -0.285(2)\,{\rm GeV}^{-2}
\end{eqnarray}
for $\beta=2.4$ and
\begin{eqnarray}
\beta_1&=&0.395(4)\,{\rm GeV}\nonumber\\
\lambda_1&=& -0.336(3)\,{\rm GeV}^{-2}
\end{eqnarray}
for $\beta=2.8$.

\begin{table}
\caption{Listed here are the results of the interpolated form factors $E_1(0)$, $M_2(0)$, and
$E_3(0)$, as well as the transition widths.  The continuum limits are also given. All the results
are in physical units. The widths can be compared with the PDG data $\Gamma=380(30)\,{\rm
keV}$~\cite{PDG:2012}.}\label{tab_2pp}
\begin{ruledtabular}
\begin{tabular}{ccccc}
$\beta$ & $E_1(\textrm{GeV})$ & $M_2(\textrm{GeV})$ & $E_3(\textrm{GeV})$  & $\Gamma(\textrm{keV})$\\
    \hline
 $2.4$    & $2.04(2)$  & $-0.218(4)$   & $0.014(3)$    & $347\pm 20$   \\
 $2.8$    & $2.08(2)$  & $-0.171(10)$  & $0.005(8)$    & $352\pm 11$   \\
 Cont.    & $2.11(2)$  & $-0.141(15)$  & $-0.007(12)$   & $361\pm 9$
\end{tabular}
\end{ruledtabular}
\end{table}


Using the interpolated form factors $F_k(0)$ and taking the fine structure constant $\alpha=1/137$,
the transition width can be calculated directly. As shown in Table~\ref{tab_2pp}, the partial decay
width of $\Gamma(\chi_{c2} \to J/\psi\ \gamma)$ is predicted to be $347\pm20$ keV or $352\pm11$ keV
for the two lattices, respectively. The continuum extrapolation gives $\Gamma=361\pm 9\,{\rm keV}$.
All these results can be compared with the PDG average of $380(30)\,{\rm keV}$. The agreement with
experimental data of the $\chi_{c2} \to J/\psi\ \gamma$ transition indicates that our method for
the $\chi_{c2}\rightarrow J/\psi$ transition is reliable. Then we can turn to transition the
$\eta_{c2}\rightarrow \gamma J/\psi$.

\subsection{$\eta_{c2} \to J/\psi\gamma$ transition}
\begin{figure}[tbh!]
\includegraphics[scale=0.6]{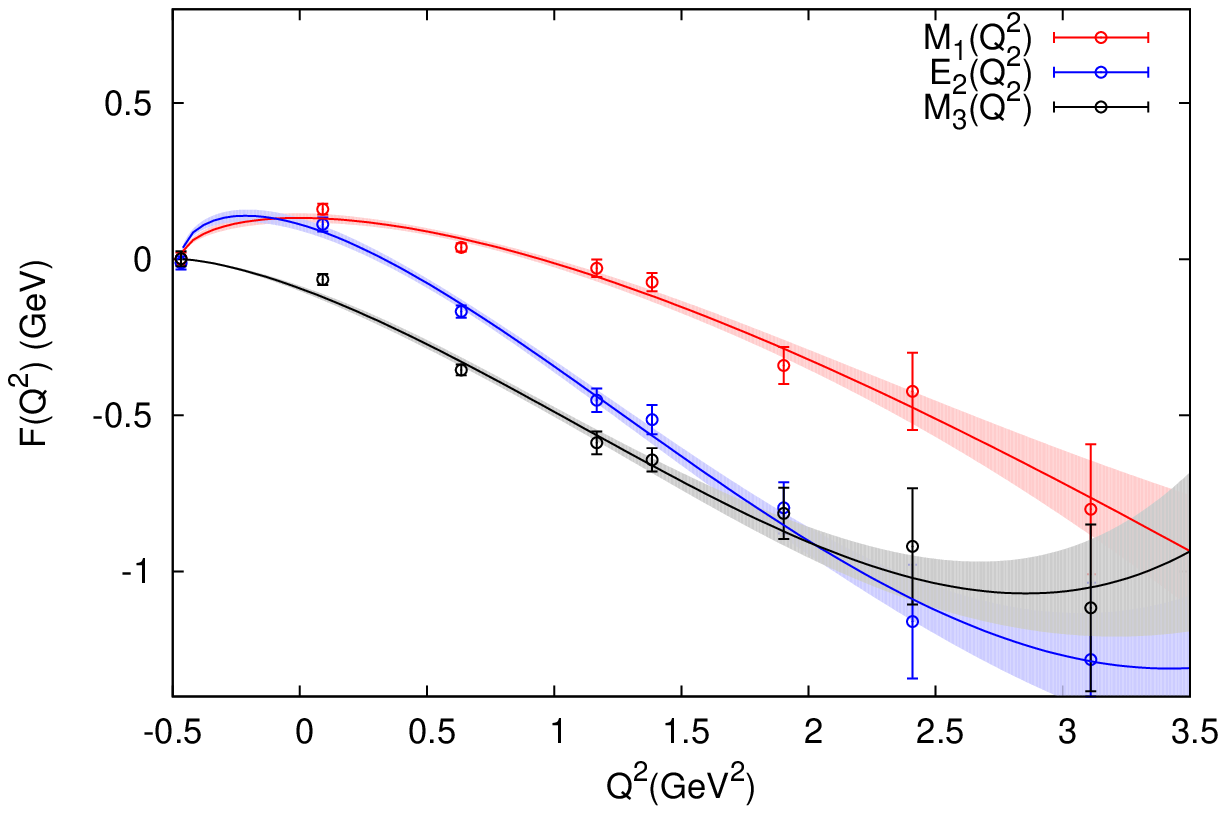}
\includegraphics[scale=0.6]{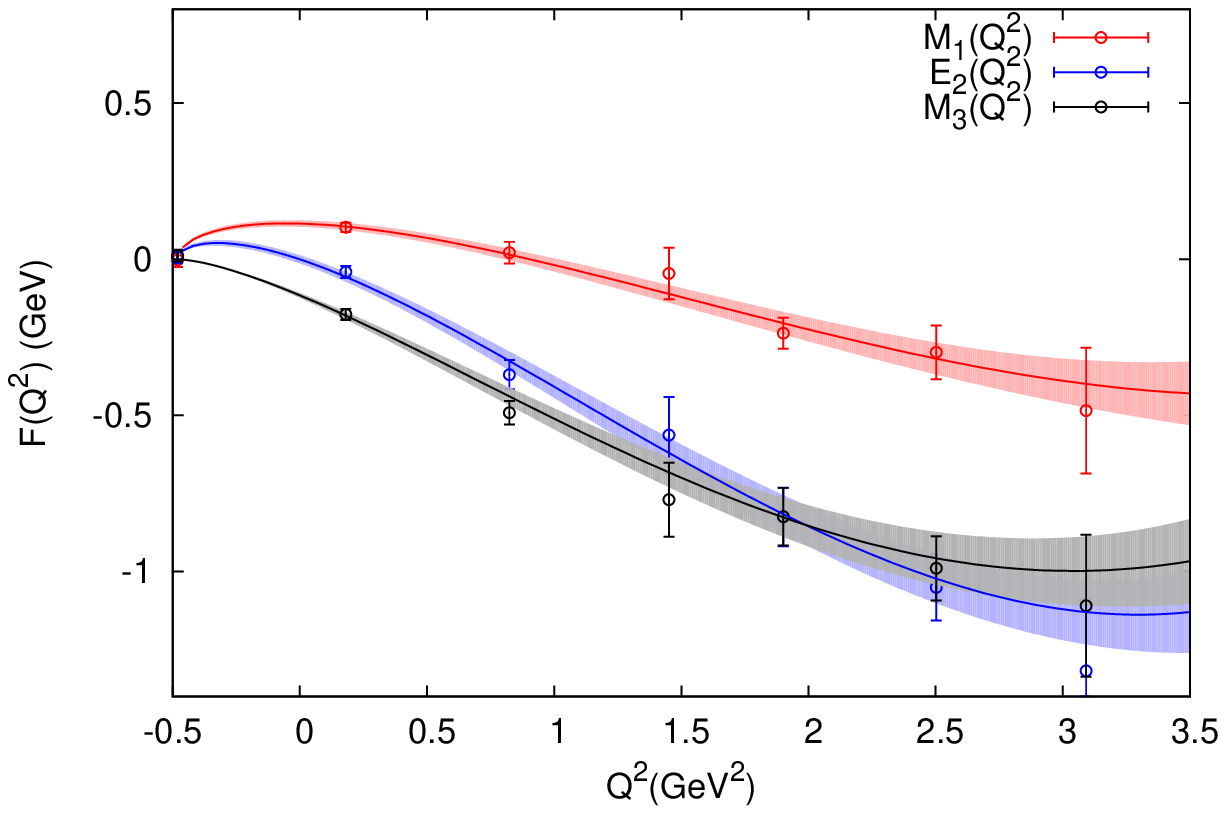}
\caption{\label{fig_2mp} The $\eta_{c2}-J/\psi$ transition form factors $M_1(Q^2)$, $E_2(Q^2)$, and
$M_3(Q^2)$ are plotted versus $Q^2$ for the two lattices of $\beta=2.4$ (the upper panel) and
$\beta=2.8$ (the lower panel), respectively. The points are the simulation data, and the lines
illustrate the fit function with jackknife error bands.}
\end{figure}

The general Lorentz decomposition of the Minkowski matrix elements responsible for the transition
$\eta_{c2} \to \gamma J/\psi$ can be expressed as
\begin{eqnarray}\label{2-+deco1}
 &&\langle V(\vec{p}_V, \lambda_V) | j^{\mu}(Q^2) | T(\vec{p}_T, \lambda_T) \rangle =
 {a(Q^2)}A^{\mu}\nonumber\\
 &&+{b(Q^2)}B^{\mu}+{c(Q^2)}C^{\mu}+{d(Q^2)}D^{\mu}+{e(Q^2)}E^{\mu}
\end{eqnarray}
where $T$ stands now for the tensor meson $\eta_{c2}$;  $a(Q^2)$, $b(Q^2)$, $c(Q^2)$, $d(Q^2)$, and
$e(Q^2)$ are Lorentz-invariant scalar functions of $Q^2$, and $A^\mu$, $B^\mu$, $C^\mu$, $D^\mu$
are kinematic functions whose explicit expressions can be found in Appendix~\ref{sec_a_pole2mp}.
With the multipole decomposition, the matrix elements can be also expressed in terms of form
factors $M_1$, $E_2$, $M_3$, and $C_2$:
\begin{eqnarray}\label{eq_mTV}
&&\langle V(\vec{p}_V, \lambda_V) | j^{\mu}(0) |T(\vec{p}_T, \lambda_T) \rangle = i\alpha_1^\mu {\
M_1(Q^2)}\nonumber\\ &&+ i\alpha_2^\mu {\ E_2(Q^2)} + i\alpha_3^\mu {\ M_3(Q^2)} -i \alpha_4^\mu {\
C_2(Q^2)}
\end{eqnarray}
where $\alpha_i^\mu$ are also kinematic functions which can be expressed in terms of the kinematic
functions in Eq.~(\ref{2-+deco1}). (See Appendix~\ref{sec_a_pole2mp}.) With real photons in the
transition $\eta_{c2} \to \gamma J/\psi$, only three multipoles are contributing: the magnetic
dipole ($M_1$), the electric quadrupole ($E_2$), and $M_3$. The transition width is written as
\begin{equation}\label{width_2mp}
\Gamma(\eta_{c2}\rightarrow \gamma
J/\psi)=\frac{16\alpha|\vec{q}|}{45M^2_{\eta_{c2}}}(|M_1(0)|^2+|E_2(0)|^2+|M_3(0)|^2).
\end{equation}
Since they are calculated at $Q^2\neq 0$, the multipole amplitudes should be interpolated to
$Q^2=0$. The form factor $C_2(Q^2)$ corresponds to the emission of longitudinal photons and does
not contribute at $Q^2=0$. In extracting the amplitudes, we take the standard procedure as
described in Sec. II. The three-point functions are calculated by setting the tensor at rest and
making the vector moving. In analogy with the $2^{++}$ case, the effect of rotational symmetry
breaking between $T_2$ irreps and $E$ irreps is found to be small in this case and is neglected in
the data analysis. The form factors $M_1(Q^2)$, $E_2(Q^2)$, and $M_3(Q^2)$ with various $Q^2$ are
extracted jointly by a correlated fitting with a one-eliminating jackknife covariance matrix, and
the results are illustrated in Fig.~\ref{fig_2mp} as data points with jackknife errors. In the
following discussion, we will focus on the interpolation procedure. What is interesting is the
relation between the two sets of form factors. We should first mention that the form factors, given
the fact that they are functions of $Q^2$, can also be written in terms of another Lorentz
invariant variable, $\Omega$:
\begin{eqnarray}
\Omega &\equiv& (p_V\cdot p_T)^2-m_V^2 m_T^2\nonumber\\
&=&\frac{1}{4}[(m_V+m_T)^2+Q^2][(m_V-m_T)^2+Q^2].
\end{eqnarray}
Thus the two sets of the form factors are related to each other as follows,
\begin{widetext}
\bea \label{relation}M_1(\Omega)&=&i\frac{\sqrt{\Omega}}{m_T} \left[\sqrt{\frac{5}{12}}\left(a\
m_V+a\ m_T-2 c\ m_T\right)+\frac{2 a\ m_V-3 c\  m_T-4 d\  m_V^2 m_T+6 e\  m_V m_T^2 }{4
\sqrt{15}}\left(\frac{\Omega}{ m_V^2 m_T^2}\right)\right.\nonumber\\
&& +\left.\frac{-2 a\ m_V+3 c\ m_T }{16 \sqrt{15}}\left(\frac{\Omega}{m_T^2 m_V^2}\right)^2
+O\left(\left(\frac{\Omega}{m_T^2 m_V^2}\right)^3\right)\right],
\nonumber\\
E_2(\Omega)&=&i\frac{\sqrt{\Omega}}{m_T}\left[-\sqrt{\frac{3}{4}} \left(a\ m_T-a\ m_V
\right)+\frac{2 a\ m_V-c\ m_T- 4 d\ m_V^2 m_T+2 e\  m_V m_T^2}{4 \sqrt{3}}\left(\frac{\Omega}{
m_V^2 m_T^2}\right)\right.\nonumber\\&&
+\left.\frac{-2 a\ m_V+c\  m_T }{16 \sqrt{3}}\left(\frac{\Omega}{m_T^2 m_V^2}\right)^2+
O\left(\left(\frac{\Omega}{m_T^2 m_V^2}\right)^3\right)\right],\nonumber\\
M_3(\Omega)&=&i\frac{\sqrt{\Omega}}{m_T}\left[-\frac{-a\ m_V-c\  m_T+2 d\  m_V^2 m_T+ 2 e\  m_V
m_T^2}{\sqrt{15}}\left(\frac{\Omega}{ m_V^2 m_T^2}\right)-\frac{a\ m_V+ c\  m_T}{4
\sqrt{15}}\left(\frac{\Omega}{m_T^2 m_V^2}\right)^2 \right.\nonumber\\
&&+\left.O\left(\left(\frac{\Omega}{m_T^2m_V^2}\right)^3\right)\right]. \eea
\end{widetext}
It is seen that each multipole form factor can be expressed as a series of $\Omega/(m_V^2m_T^2)$
with a prefactor $\sqrt{\Omega}/m_T$. In the rest frame of $T$ (as is the case in our calculation),
the expression of $\Omega$ is simplified as $\Omega=(m_T |\vec{p}_V|)^2$, such that
$\Omega/(m_V^2m_T^2)=v^2$, with $v=|\vec{p}_V|/m_V$ being the spatial velocity of $V$. The
convergence of the series in $v$ is guaranteed if the form factors $a$, $b$, $c$, $d$, $e$ are not
singular in $Q^2$, since $v<1$ (For our calculation, the largest value of $v$ is approximately
$0.5$). So, for the decaying $T$ at rest, we have the simplified expression of the form factors
$M_1$, $E_2$, and $M_3$:
\begin{eqnarray}\label{relation2}
M_1&=&|\vec{p}_V|(A_1(Q^2)+B_1(Q^2)v^2+C_1(Q^2)v^4+O(v^6))\nonumber\\
E_2&=&|\vec{p}_V|(A_2(Q^2)+B_2(Q^2)v^2+C_2(Q^2)v^4+O(v^6))\nonumber\\
M_3&=&|\vec{p}_V|(B_3(Q^2)v^2+C_3(Q^2)v^4+O(v^6)).
\end{eqnarray}
With these expressions, the following information can be inferred: (i) The desired $|\vec{p}_V|$
prefactor accounting for the $P$-wave decay of $\eta_{c2}\to \gamma J/\psi$ is explicitly derived.
(ii) The leading contribution to $M_1$ and $E_2$ is of order $O(1)$, while that of $M_3$ is of
order $O(v^2)$. For the case of this study, since $v_{\rm max}\sim 0.5$, it is reasonable that the
nonsingular $A_i,B_i$ and $C_i$ can be expanded with respect to $v$, such that we can take the
following functions to do the interpolation:
\begin{eqnarray}\label{fit_2mp}
F_i(v)&=& Av+Bv^3+Cv^5+O(v^6)(F_i\to M_1,~E_2)\nonumber\\
F_i(v)&=& Bv^3+Cv^5+Dv^7 +O(v^9)(F_i\to M_3),
\end{eqnarray}
and the on-shell amplitudes $M_1(Q^2=0)$, $E_2(Q^2=0)$, and $M_3(Q^2=0)$ can be reached by
$F_i(v_0)$ with $v_0=(m_T^2-m_V^2)/(2m_Tm_V)$. The extracted form factor and the interpolation are
 shown in Fig.~\ref{fig_2mp}, where the data points are the simulated results with jackknife errors.
 One can see that at $v=0$ [corresponding to $Q^2=-(m_T-m_V)^2\sim 0.5\,{\rm GeV}^2$] the form
 factors $M_1$, $E_2$, and $M_3$ are surely consistent with zero. The fits using
 Eq.~(\ref{fit_2mp})
 are also shown as curves with jackknife error bands. The interpolated values of these form factors
 at $Q^2=0$ for both $\beta=2.4$ and $\beta=2.8$ are listed in Table~\ref{tab_2mp}, where the
 resultant transition widths and the corresponding continuum limits are also given. It is surprising
 that, for both lattices, the obtained $|M_3|$ is unexpectedly large and comparable to $M_1$. This
 may be qualitatively attributed to recoiling effects of the charm quark or charm antiquark by emitting the hard
 photon with an energy $E_\gamma\sim 0.6\,{\rm GeV}$ in this transition, which may result in large
 form factors $d(Q^2)$ and $e(Q^2)$ (see the discussion below). In contrast to the mild dependence of $M_1$ and
 $M_3$ on the lattice spacing, the form factor $E_2$ is very sensitive to the lattice spacing.
 The reason for this is unclear and under investigation. Anyway, after a naive continuum
 extrapolation using the data from the two lattices in this work, we get the continuum results of
 the form factors as follows:
\begin{eqnarray}\label{eq_formus}
M_1&=&0.104(10)\textrm{GeV},\nonumber\\
E_2&=&-0.071(20)\textrm{GeV},\nonumber\\
M_3&=&-0.132(10)\textrm{GeV}.
\end{eqnarray}
Applying these results to Eq.~(\ref{width_2mp}), the transition width of $\eta_{c2}\to \gamma
J/\psi$ is predicted to be
\begin{equation}
\Gamma(\eta_{c2}\rightarrow \gamma J/\psi)=3.8\pm 0.9\,{\rm keV}.
\end{equation}
\begin{table}[t]
\caption{\label{tab_2mp} Listed here are the interpolated values of the form factors $M_1$, $E_2$,
and $M_3$ at $Q^2=0$ for both $\beta=2.4$ and $\beta=2.8$. The
 resultant transition widths and the corresponding continuum limits are also given.}
\begin{ruledtabular}
\begin{tabular}{ccccc}
$\beta$ & $M_1(\textrm{GeV})$ & $E_2(\textrm{GeV})$ & $M_3(\textrm{GeV})$  & $\Gamma(\textrm{keV})$\\
    \hline
 $2.4$    & $0.133(13)$  & $0.111(17)$   & $-0.093(9)$    & $4.4\pm0.9$   \\
 $2.8$    & $0.115(11)$  & $-0.0007(14)$ & $-0.117(9)$    & $3.1\pm0.6$   \\
 Cont.    & $0.104(10)$  & $-0.071(20)$  & $-0.132(10)$   & $3.8\pm0.9$
\end{tabular}
\end{ruledtabular}
\end{table}
There have also been several phenomenological studies on this transition, one of which is in the
framework of the light-front quark model~\cite{Ke:2011jf}, where the on-shell transition amplitude
is decomposited as
\bea
&&\langle V(\vec{p}_V, \lambda_V) | j^{\mu}(0) | \eta_{2}(\vec{p}_T, \lambda_T) \rangle = \nonumber \\
&&\Big[2f_1\epsl^{\mu\nu\rho\sigma}p^T_{\nu} p^V_{\rho}\epsilon^{\beta}_{\sigma}(\vec{p}_T,
\lambda_T) \epsilon^*_{\beta}(\vec{p}_V, \lambda_V)
\nonumber\\
&&+(f_2+f_3)\epsl^{\mu\nu\rho\sigma}p^T_{\nu} p^V_{\rho}\epsilon^*_{\sigma}(\vec{p}_V, \lambda_V)
\epsilon^{\alpha \beta}(\vec{p}_T, \lambda_T)p^V_{\alpha}p^V_{\beta}\ \nonumber \\&&
+2f_4\epsl^{\mu\nu\rho\sigma}p^T_{\nu} p^V_{\rho}\epsilon^{\beta}_{\sigma}(\vec{p}_T,
\lambda_T)p^V_{\beta} \epsilon^{*\alpha}(\vec{p}_V, \lambda_V)p^T_{\alpha}\Big],\eea
and the effective couplings are determined to be
\begin{eqnarray}\label{eq_lifn}
f_1&=&-0.0140(2)\textrm{GeV}^{-1},\nonumber\\
f_2&=& 0.146(3)\textrm{GeV}^{-3},\nonumber\\
f_3&=&-0.092(1)\textrm{GeV}^{-3},\nonumber\\
f_4&=&0.0180(1)\textrm{GeV}^{-3}.
\end{eqnarray}
Since this decomposition is equivalent to Eq.~(\ref{2-+deco1}) by the relation  \bea c(0)=2f_1,\
d(0)=-(f_2+f_3),\ e(0)=-2f_4 \eea  [It should be notified that $a(Q^2)$ and $b(Q^2)$ are equal to
zero when $Q^2=0$ because they are proportion to $C_2(Q^2)$.], the corresponding multipole
amplitudes can be calculated from Eq.~(\ref{relation}) as
\begin{eqnarray}
\label{eq_licasea}
M_1&=&0.079(2)\textrm{GeV},\nonumber\\
E_2&=&-0.086(2)\textrm{GeV},\nonumber\\
M_3&=&-0.125(3)\textrm{GeV},
\end{eqnarray}
which gives a width of $\Gamma$=3.54(12) keV. Taking into consideration the uncertainty of the
choice of parameters such as the charm quark mass $m_c$ and the wave function parameter, etc., one
can find that the lattice results and the LFQM results are surprisingly in excellent agreement. On
the other hand, with the values in Eq.~(\ref{eq_lifn}), we find that the coefficients $B_i(Q^2)$ of
the $v^2$ term in Eq.~(\ref{relation2}) are surely much larger than $A_i(Q^2)$ at $Q^2=0$ so as to
compensate for the suppression of $v^2$. This explains to some extent the fact that the $M_3$ in
this transition competes $M_1$ and $E_2$.

The other phenomenological study~\cite{Jia:2010jn} applying the nonrelativistic QCD (NRQCD) gives
the transition width as
\begin{equation} \Gamma(\eta_{c2}\to \gamma
J/\psi)=\frac{8\alpha|\vec{k}|^3}{675m^2_c}(a_1^2+a_2^2+a_3^2)\label{eq_mTV_ph}
\end{equation}
where $a_1$, $a_2$, and $a_3$ are equivalent to the standard formfactors $M_1$, $E_2$, and $M_3$ up
to a constant factor and are calculated explicitly in NRQCD. By comparing this equation with
Eq.~(\ref{eq_form}), the factor is approximately $0.33\,{\rm GeV}$, say, $F_i\simeq (0.33\,{\rm
GeV}) a_i$. Thus, their work gives the predictions
\begin{equation}
M_1\sim 0.026-0.045\,{\rm GeV} ,E_2\simeq M_3 \simeq -0.13\,{\rm GeV},
\end{equation}
which are also in reasonable agreement with our results.

\section{Conclusion}\label{section4}
We calculate in the quenched approximation the mass of $J^{PC}=2^{-+}$ charmonium $\eta_{c2}$, as
well as its radiative transition width to $J/\psi$. The computations are carried out on two
anisotropic lattices with different lattice spacings, such that the lattice artifacts can be
controlled to some extent. As a calibration, we calculate first the spectrum of the lowest-lying
charmonia, such as $1S$ and $1P$ states, and reproduce the physical pattern of the spectrum. In
addition, we calculate the transition width of $\chi_{c2}\rightarrow \gamma J/\psi$ and get the
result $361\pm 9\,{\rm keV}$, which is in good agreement with the experimental value of $380\pm
30\,{\rm keV}$. Both of these facts manifest the small systematic uncertainties due to the quenched
approximation and the finite lattice spacings.

There are two states observed in the $2^{-+}$ channel, with masses $3.80(3)\,{\rm GeV}$ and
$4.43(8)\,{\rm GeV}$. The lower state has a mass similar to that of the well-established
$\psi(3770)$, which is always assigned to be mainly the $1{}^3D_1$ charmonium, and therefore can be
naturally identified as the conventional $1{}^1D_2$ charmonium $\eta_{c2}$. Obviously, it is about
70 MeV lower in mass than $X(3872)$, and this difference cannot be attributed to the systematic
uncertainties of our work.

As for the transition rate of $\eta_{c2} \to J/\psi\gamma$, we get a small partial width of roughly
3.8(9) keV, which is in agreement with the phenomenological studies. Taking the branch ratio ${\rm
Br}(X(3872)\rightarrow J/\psi \gamma)> 0.9\%~(BABAR)$ or $0.6\%~({\rm Belle})$, the full width of
$X(3872)$ is estimated to be $<420-630$ keV, which is smaller than, but not in contradiction with
the experimental upper limit $\Gamma_X<1.2\,{\rm MeV}$. Obviously, a reliable calculation of the
partial width $\eta_{c2}\to \psi' \gamma $ is also crucial for the ${}^1D_2$ charmonium assignment
of $X(3872)$, but unfortunately there are difficulties in the unambiguous extraction of excited
states on the lattice. However, we can still infer some useful information from the calculation of
$\eta_{c2} \to J/\psi\ \gamma$. In the potential quark model, it is known that ${}^1D_2\to \gamma
V$ is a hindered transition with $M_1(0)=0$; therefore, the observed nonzero $M_1(0)$ and the
appearance of the higher multipoles $E_2$ and $M_3$ can be understood as the relativistic
correction and the recoil effects of the emission of a hard photon ($E_\gamma\sim 0.65\,{\rm GeV}$
for the final $J/\psi$ and $E_\gamma\sim 0.11\,{\rm GeV}$ for $\psi'$). Intuitively, this kind of
effect for the final $\psi'$ can be similar to that for the final $J/\psi$, or even milder; thus,
the width of $\eta_{c2}\to \gamma \psi'$ will be suppressed by a kinematic factor of
$(0.65/0.11)^3\sim 200$ when compared with the transition $\eta_{c2}\to \gamma J/\psi$. With this
fact in mind, the $\eta_{c2}$ assignment of $X(3872)$ can be ruled out if {\it BABAR}'s observation
of ${\rm Br}(X(3872)\to \gamma \psi')/{\rm Br}(X(3872)\to \gamma J/\psi)=3.4\pm 1.4$ is confirmed.

\section*{ACKNOWLEDGEMENTS}

This work was supported in part by the National Science Foundation of China (NSFC) under Grants No.
10835002, No. 11075167, No. 11021092, No. 10975076, and No. 11105153. Y. Chen and C. Liu also thank
the support of NSFC and DFG (CRC110).

\appendix

\section{Multipoles decomposition of $\chi_{c2}\leftrightarrow J/\psi$}\label{sec_a_pole2pp}
For the convenience of readers, the details of the multipole decomposition of matrix elements of
the electromagnetic current $j^{\mu}(0)$ between a $1^{--}$ vector $V$ state and a $2^{++}$ tensor
state are described here following Ref.~\cite{Dudek2009}. The most general Lorentz-covariant
decomposition with $P$ and $C$ parity invariance is
\begin{eqnarray}\label{general_tensor}
&&\langle V(\vec{p}_V, \lambda_V) | j^{\mu}(0) | T(\vec{p}_T, \lambda_T)\rangle = a(Q^2) A^{\mu}
+ b(Q^2)  B^{\mu} \nonumber\\
&&~~~~~~+ c(Q^2) C^{\mu}+ d_T(Q^2)  D_T^{\mu} + d_V(Q^2)  D_V^{\mu}\nonumber\\
&&~~~~~~+ f_T(Q^2) F_T^{\mu} + f_V(Q^2) F_V^{\mu}
\end{eqnarray}
with the definitions
 \begin{eqnarray}\label{eq_pole2pp}
A^{\mu} &\equiv& \epsilon^{\mu \nu}(\vec{p}_T, \lambda_T) \epsilon^*_{\nu}(\vec{p}_V,\lambda_V);\nonumber\\
B^{\mu} &\equiv& \epsilon^{\mu \nu}(\vec{p}_T, \lambda_T) p^V_{\nu} (\epsilon^*(\vec{p}_V, \lambda_V) \cdot p_T);\nonumber \\
C^{\mu} &\equiv& \epsilon^{*\mu}(\vec{p}_V, \lambda_V) (\epsilon^{\alpha \beta}(\vec{p}_T, \lambda_T) p^V_{\alpha} p^V_{\beta}); \nonumber\\
D_T^{\mu} &\equiv& p_T^{\mu} (\epsilon^{\alpha \beta}(\vec{p}_T, \lambda_T) \epsilon^*_{\alpha}(\vec{p}_V, \lambda_V) p^V_{\beta});\nonumber \\
D_V^{\mu} &\equiv& p_V^{\mu} (\epsilon^{\alpha \beta}(\vec{p}_T, \lambda_T) \epsilon^*_{\alpha}(\vec{p}_V, \lambda_V) p^V_{\beta});\nonumber \\
F_T^{\mu} &\equiv& p_T^{\mu} (\epsilon^{\alpha \beta}(\vec{p}_T, \lambda_T) p^V_{\alpha} p^V_{\beta})
(\epsilon^*(\vec{p}_V, \lambda_V) \cdot p_T);\nonumber \\
F_V^{\mu} &\equiv& p_V^{\mu} (\epsilon^{\alpha \beta}(\vec{p}_T, \lambda_T) p^V_{\alpha}
p^V_{\beta}) (\epsilon^*(\vec{p}_V, \lambda_V) \cdot p_T).  \nonumber\\
\end{eqnarray}
On the other hand, the matrix elements $\langle V|j^\mu|T\rangle$ can be also expressed in terms of
the helicity amplitudes:
\begin{widetext}
\bea
  \langle V | j^{\mu} | T \rangle \epsilon^*_{\mu}(\lambda_{\gamma}=\pm) &=&   \sum_k \sqrt{\frac{2k+1}{2J+1}}
   \left[ E_k \tfrac{1}{2}(1+ (-1)^k \delta P ) \mp M_k
     \tfrac{1}{2}(1- (-1)^k \delta P )\right] \langle k\mp;
   J'\lambda\pm1 | J\lambda\rangle \label{eq_helipm}\\
 \langle V | j^{\mu} | T \rangle \epsilon^*_{\mu}(\lambda_{\gamma}=0) &=&
 \sum_k \sqrt{\frac{2k+1}{2J+1}}  C_k \tfrac{1}{2}(1+ (-1)^k \delta P ) \langle k0; J'\lambda | J\lambda\rangle
 \label{eq_heli0},
\eea
\end{widetext}
where $E_k$, $M_k$, and $C_K$ are multipole amplitudes, and $\delta P$ is the product of the $P$
parity of the initial ($T$) and finial ($V$) states, taking the value $\delta P=-1$ for the
$2^+\rightarrow 1^-$ transition. These are actually the helicity selection rules. An additional
constraint comes from the conservation of the vector current,
\begin{equation}\label{eq_curr_con}
\langle V | j^{\mu} | T \rangle q_{\mu}= \partial_{\mu}\langle V | j^{\mu} | T \rangle = 0.
\end{equation}
With the constraints of Eqs.~(\ref{eq_helipm},~\ref{eq_heli0},~\ref{eq_curr_con}), we can solve
$a(Q^2)$, $b(Q^2)$, $c(Q^2), \ldots$ in the rest frame of the initial state with the spatial
momentum of the photon parallel to the $z$ axis. [The polarization vector of the photon takes
$(1,0,0,1)$.] Thus, we can get the expressions in terms of $E_k(q^2)$, $C_K(Q^2)$. After that, the
general expression of the form factor $a$, $b$, $c$, $\ldots$ can be obtained by carrying out a
general Lorentz transformation. For the case of the $2^+\rightarrow 1^-$ transition here,
Eq.~(\ref{eq_helipm}) provides three independent equations with respect to the three different
helicities of the vector state, and therefore gives the relations between [$a(Q^2),b(Q^2),c(Q^2)$]
and $E_1(Q^2)$, $M_2(Q^2)$, $E_3(Q^2)$] as
\begin{eqnarray}\label{eq_chi_abc}
a&=&\frac{E_3 }{\sqrt{15}}-M_2 {\sqrt{3}}+\sqrt{\frac{3}{5}} E_1,\nonumber\\
b&=& \frac{1}{5 \sqrt{3} \Omega }\left(3 \sqrt{5} E_1 (m_T m_V-p_T.p_V)\right.\nonumber\\
&& + 5 M_2 (m_I m_V+p_T.p_V)\nonumber \\
&& -\left. \sqrt{5} E_3 (4 m_T m_V+p_T.p_V)\right)\nonumber\\
c&=& m_T^2\frac{\sqrt{5} E_3+4M_2}{2 \sqrt{3} \Omega },
\end{eqnarray}
with $\Omega\equiv(p_T\cdot p_V)^2-m_T^2m_V^2$. The constraints from Eqs.~(\ref{eq_heli0})
and~(\ref{eq_curr_con}) with plus/minus helicity of the vector meson can fix the parameters $d_V$
and $d_T$. Furthermore, $f_V$ and $f_T$ can be derived from Eqs.~(\ref{eq_heli0})
and~(\ref{eq_curr_con}) with zero helicity of the vector meson. As such, the multipole
decomposition of the matrix elements $\langle V|j^\mu|T\rangle$ can be expressed finally as
\begin{eqnarray}\label{multipole_2++}
&&\langle V(\vec{p}_V, \lambda_V) | j^{\mu}(Q^2) | T(\vec{p}_T, \lambda_T)\rangle = \alpha_1^\mu
E_1(Q^2)\nonumber\\
&&+  \alpha_2^{\mu}M_2(Q^2)+\alpha_3^\mu E_3(Q^2) + \alpha_4^\mu C_1(Q^2)+ \alpha_5^\mu C_3(Q^2)
\nonumber\\
\end{eqnarray}
with the functions $\alpha_i^\mu$ defined by
\begin{widetext}
\begin{eqnarray}\label{eq_TV}
\alpha_1^\mu &=& \sqrt{\frac{3}{5}} \Bigg[ -A^{\mu} + \frac{m_T}{\Omega}(\tilde{\omega}-m_V)B^{\mu}
+ \frac{m_T}{\Omega}\Big( \tilde{\omega} D_T^{\mu} - m_T D_V^{\mu} \Big) +
\frac{m_T^2}{\Omega^2}(\tilde{\omega}-m_V)
\Big( -\tilde{\omega} F_T^{\mu} + m_T F_V^{\mu} \Big)  \Bigg]\nonumber \\
\alpha_2^\mu &=& \sqrt{\frac{1}{3}} \Bigg[ A^{\mu} - \frac{m_T}{\Omega}(\tilde{\omega}+m_V)B^{\mu}
- \frac{2 m_T^2}{\Omega} C^{\mu} + \frac{m_T}{\Omega}\Big( -\tilde{\omega} D_T^{\mu} + m_T D_V^{\mu} \Big)\nonumber \\
&&~~~~~~+ \frac{m_T^2}{\Omega^2}\Big( (\tilde{\omega}^2 +
\tilde{\omega} m_V - 2m_V^2) F_T^{\mu} + m_i(\tilde{\omega}-m_V)F_V^{\mu} \Big)  \Bigg]\nonumber \\
\alpha_3^\mu &=&  \sqrt{\frac{1}{15}} \Bigg[ -A^{\mu} +
\frac{m_T}{\Omega}(\tilde{\omega}+4m_V)B^{\mu}
- \frac{5 m_T^2}{2\Omega} C^{\mu} + \frac{m_T}{\Omega}\Big( \tilde{\omega} D_T^{\mu} - m_T D_V^{\mu} \Big) \nonumber\\
&&~~~~~~+ \frac{m_T^2}{\Omega^2}\Big( -(\tilde{\omega}^2 + 4 \tilde{\omega} m_V +
\tfrac{5}{2}m_V^2) F_T^{\mu}
+ m_i(\tfrac{7}{2}\tilde{\omega}+4m_V)F_V^{\mu} \Big)  \Bigg] \nonumber\\
\alpha_4^\mu &=& \sqrt{\frac{3}{5}}\frac{m_T}{\Omega \sqrt{q^2}} \Bigg[
 (m_V^2-\tilde{\omega} m_T ) D_T^{\mu} + (m_T^2 - \tilde{\omega} m_T ) D_V^{\mu}-\frac{m_T}{\Omega}\big(\tilde{\omega}
 -m_V\big) \Big((m_V^2-\tilde{\omega} m_T ) F_T^{\mu}
+ (m_T^2-\tilde{\omega}  m_T ) F_V^{\mu} \Big) \Bigg]\nonumber \\
\alpha_5^\mu &=& \sqrt{\frac{2}{5}}\frac{m_T}{\Omega \sqrt{q^2}} \Bigg[
 (m_V^2-\tilde{\omega} m_T ) D_T^{\mu} + (m_T^2-\tilde{\omega} m_T ) D_V^{\mu}-\frac{m_T}{\Omega}\big(\tilde{\omega} +
 \tfrac{3}{2}m_V\big) \Big((m_V^2-\tilde{\omega} m_T )
F_T^{\mu} + (m_T^2-\tilde{\omega}  m_T ) F_V^{\mu} \Big) \Bigg].\nonumber\\
\end{eqnarray}
\end{widetext}
where $\tilde{\omega}\equiv \frac{p_T.p_V}{m_T}$ and $\omega^{\pm} \equiv p_T.p_V\pm m_Vm_T$.

\section{Multipoles of $\eta_{c2}\leftrightarrow J/\psi$}\label{sec_a_pole2mp}

The case of $\eta_{c2}\rightarrow J/\psi$ is sightly different from that above. The most general
Lorentz-covariant decomposition with $P$ and $C$ parity invariance is
\begin{eqnarray}\label{general_2-+}
&&\langle V(\vec{p}_V, \lambda_V) | j^{\mu}(0) | \eta_{2}(\vec{p}_T, \lambda_T) \rangle =
 a(Q^2)A^{\mu}+b(Q^2)B^{\mu}\nonumber\\
 &&~~~~~~~~+c(Q^2)C^{\mu}+d(Q^2)D^{\mu}+e(Q^2)E^{\mu}
\end{eqnarray}
with
\begin{eqnarray}\label{independant}
A^{\mu}&=&\epsl^{\mu\nu\rho\sigma}\epsilon^*_{\nu}(\vec{p}_V, \lambda_V)
p^V_{\rho}\epsilon^{\beta}_{\sigma}(\vec{p}_T, \lambda_T)p^V_{\beta},\nonumber\\
B^{\mu}&=&\epsl^{\beta\nu\rho\sigma}p^T_{\beta}\epsilon^*_{\nu}(\vec{p}_V, \lambda_V)
p^V_{\rho}\epsilon^{\mu}_{\sigma}(\vec{p}_T, \lambda_T),\nonumber\\
C^{\mu}&=&\epsl^{\mu\nu\rho\sigma}p^T_{\nu} p^V_{\rho}\epsilon^{\beta}_{\sigma}(\vec{p}_T,
\lambda_T)\epsilon^*_{\beta}(\vec{p}_V, \lambda_V),\ \nonumber \\
D^{\mu}&=&\epsl^{\mu\nu\rho\sigma}p^V_{\nu} p^T_{\rho}\epsilon^*_{\sigma}(\vec{p}_V,
\lambda_V)\epsilon^{\alpha \beta}(\vec{p}_T, \lambda_T)p^V_{\alpha}p^V_{\beta},\ \nonumber \\
E^{\mu}&=&\epsl^{\mu\nu\rho\sigma}p^V_{\nu} p^T_{\rho}\epsilon^{\beta}_{\sigma}(\vec{p}_T,
\lambda_T)p^V_{\beta}\epsilon^{*\alpha}(\vec{p}_V, \lambda_V)p^T_{\alpha}.\
\end{eqnarray}
In fact, there exist another three Lorentz-covariant structures $A_T$, $E_V$, and $E_T$:
\begin{eqnarray}
E_V^{\mu}&=&\epsl^{\alpha\nu\rho\sigma}p^T_{\alpha}\epsilon^*_{\beta}(\vec{p}_V, \lambda_V)
p^V_{\rho}\epsilon^{\sigma \beta}(\vec{p}_T, \lambda_T)p^V_{\beta}p^{\mu}_V,\ \nonumber \\
A_T^{\mu}&=&\epsl^{\mu\nu\rho\sigma}\epsilon^*_{\nu}(\vec{p}_V, \lambda_V)
p^T_{\rho}\epsilon^{\beta}_{\sigma}(\vec{p}_T, \lambda_T)p^V_{\beta},\ \nonumber \\
E_T^{\mu}&=&\epsl^{\alpha\nu\rho\sigma}p^T_{\alpha}\epsilon^*_{\beta}(\vec{p}_V, \lambda_V)
p^V_{\rho}\epsilon^{\sigma \beta}(\vec{p}_T, \lambda_T)p^V_{\beta}p^{\mu}_T,
\end{eqnarray}
which, however, are not independent and can be expressed in terms of the functions in
Eq.~(\ref{independant}):
\begin{eqnarray}
A_T^{\mu}&=&-B^{\mu}-C^{\mu}\nonumber \\
E_T^{\mu}&=&m_T^2 A^{\mu}+ p_V.p_T B^{\mu}+p_V.p_T C^{\mu}+ E^{\mu}\nonumber \\
E_V^{\mu}&=&p_V\cdot p_T A^{\mu}+ m_V^2 B^{\mu}+m_V^2 C^{\mu}+ D^{\mu}.
\end{eqnarray}
So they do not appear in the decomposition. Based on this, one can follow the similar procedure of
the case of $\chi_{c2}\rightarrow J/\psi$ to derive the related multipole decomposition.  The
constraints of decomposition are similar to Eqs.~(\ref{eq_helipm},\ref{eq_curr_con}) while $\delta
P=1$. Finally, one can get the result
\begin{widetext}
\bea\label{eq_mTV} &&\langle V(\vec{p}_V, \lambda_V) | j^{\mu}(0) | \eta_2(\vec{p}_T, \lambda_T)
\rangle = \nonumber \\&& \frac{i\ M_1(Q^2)}{5\Omega^{1/2}}  \Big[-\sqrt{15} C^{\mu}
+\frac{1}{\Omega}\sqrt{15} E^{\mu}( -m_V m_T+ p_T.p_V)
 \Big] \nonumber \\&&
+\frac{i\ E_2(Q^2)}{3\Omega^{1/2}} \Big[\sqrt{3} C^{\mu}  + \frac{1}{\Omega}\Big(2 \sqrt{3} D^{\mu}
m_T^2 -\sqrt{3} E^{\mu} (m_V m_T+p_T.p_V) \Big)\Big]\nonumber \\&&
+\frac{i\ M_3(Q^2)}{30\Omega^{1/2}}\Big[- 2 \sqrt{15} C^{\mu}  + \frac{1}{\Omega}\Big(5 \sqrt{15}
D^{\mu} m_T^2  +2 \sqrt{15} E^{\mu} (4m_V m_T+p_T.p_V)\Big)\Big]\nonumber \\&&
 -\frac{i\ C_2(Q^2)}{\sqrt{q^2}\Omega^{1/2}}\Big[A^{\mu} m_T + B^{\mu} m_T + C^{\mu} m_T +
 \frac{1}{\Omega}\Big(D^{\mu} m_T(m_T^2-p_T.p_V) + E^{\mu}  m_T(m_V^2-p_T.p_V)  \Big)\Big]
\eea
\end{widetext}

\section{Form factor as function of $Q^2$ or $\Omega$}\label{sec_a_power}

Transition form factors are always expressed as Lorentz scalar functions of the squared momentum
transfer, $Q^2=-q^2=-(q_i-q_f)^2$, where $p_i$ and $p_f$ refer to the four-momenta of the initial
and final particles, respectively. However, if one looks into the Lorentz decomposition
[Eq.~(\ref{general_tensor})] and the multipole decomposition [Eq.~(\ref{multipole_2++})] for the
$\chi_{c2}\rightarrow J/\psi$ transition matrix elements, one can find that the quantity
$\Omega\equiv(p_i.p_f)^2-m_i^2m_f^2$, with $p_i.p_f=(m_i^2+m_f^2+Q^2)/2$, is also an interesting
Lorentz-invariant kinematic variable. According to Eq.~(\ref{eq_chi_abc}), the multipole amplitudes
$E_1$, $M_2$, and $E_3$ can be reversely expressed in terms of the form factors $a$, $b$, $c$ as,
\begin{widetext}
\bea E_1(\Omega)&=&\sqrt{\frac{5}{3}} a+\frac{6 c\ m_V-3 a\ m_T+4 b\  m_V^2 m_T }{4 \sqrt{15} m_T}
\left(\frac{\Omega}{m_V^2m_T^2}\right)-\frac{1}{16}\sqrt{\frac{3}{5}} a \left(\frac{\Omega}{m_V^2m_T^2} \right)^2+O\left(\left(\frac{\Omega}{m_V^2m_T^2}\right)^3\right),\nonumber\\
M_2(\Omega)&=&\frac{2 c\ m_V-a\ m_T+4 b\  m_V^2 m_T  }{4 \sqrt{3}
m_T}\left(\frac{\Omega}{m_V^2m_T^2}\right)
-\frac{a}{16 \sqrt{3}}\left(\frac{\Omega}{m_V^2m_T^2} \right)^2+O\left(\left(\frac{\Omega}{m_V^2m_T^2}\right)^3\right),\nonumber\\
E_3(\Omega)&=&\frac{-2 c\ m_V+\left(a+2 b\  m_V^2\right) m_T  }{\sqrt{15}
m_T}\left(\frac{\Omega}{m_V^2m_T^2}\right) +\frac{a}{4
\sqrt{15}}\left(\frac{\Omega}{m_V^2m_T^2}\right)^2+O\left(\left(\frac{\Omega}{m_V^2m_T^2}\right)^3\right).
\eea
\end{widetext}
Obviously, these are polynomials of the variable $\Omega/(m_V^2m_T^2)$ with the coefficients the
combinations of the form factors $a$, $b$, and $c$.  The physical meaning of the above expression
can be understood in the rest frame of the decaying particle ($T$ here), where
$\Omega/(m_V^2m_T^2)=v^2$ with $v=|\vec{p}_V|/m_V\sim 0.16$ for the $\chi_{c2}\rightarrow J/\psi$
transition. It is clearly seen from these expressions that the $E_1$ transition is dominant, while
$M_2$ and $E_3$ transitions are suppressed by a factor of $v^2\sim 0.026$. On the other hand,
$M_2(Q^2)$ and $E_3(Q^2)$ should be zero at $v=0$, or equivalently, $Q^2=-(m_T-m_V)^2$, as is
confirmed by our simulation results (seen in Fig.~\ref{fig_2pp}). It is found in the calculation
that $M_2$ and $E_3$ are consistent with zero when both the initial and final state are at rest.

As for the $\eta_{c2}\rightarrow J/\psi$ transition, the multipole amplitudes $M_1$, $E_2$ and
$M_3$ can similarly be expressed as polynomials in $\Omega/(m_V^2m_T^2)$, where the coefficients
are also the combinations of the form factors in Eq.~(\ref{general_2-+}):
\begin{widetext}
\bea M_1(\Omega)&=&i\frac{\sqrt{\Omega}}{m_T} \left[\sqrt{\frac{5}{12}}\left(a\ m_V+a\ m_T-2 c\
m_T\right)+\frac{2 a\ m_V-3 c\  m_T-4 d\  m_V^2 m_T+6 e\  m_V m_T^2 }{4
\sqrt{15}}\left(\frac{\Omega}{ m_V^2 m_T^2}\right)\right.\nonumber\\
&& +\left.\frac{-2 a\ m_V+3 c\ m_T }{16 \sqrt{15}}\left(\frac{\Omega}{m_T^2 m_V^2}\right)^2
+O\left(\left(\frac{\Omega}{m_T^2 m_V^2}\right)^3\right)\right],
\nonumber\\
E_2(\Omega)&=&i\frac{\sqrt{\Omega}}{m_T}\left[-\sqrt{\frac{3}{4}} \left(a\ m_T-a\ m_V
\right)+\frac{2 a\ m_V-c\ m_T- 4 d\ m_V^2 m_T+2 e\  m_V m_T^2}{4 \sqrt{3}}\left(\frac{\Omega}{
m_V^2 m_T^2}\right)\right.\nonumber\\&& +\left.\frac{-2 a\ m_V+c\  m_T }{16
\sqrt{3}}\left(\frac{\Omega}{m_T^2 m_V^2}\right)^2+
O\left(\left(\frac{\Omega}{m_T^2 m_V^2}\right)^3\right)\right],\nonumber\\
M_3(\Omega)&=&i\frac{\sqrt{\Omega}}{m_T}\left[-\frac{-a\ m_V-c\  m_T+2 d\  m_V^2 m_T+ 2 e\  m_V
m_T^2}{\sqrt{15}}\left(\frac{\Omega}{ m_V^2 m_T^2}\right)-\frac{a\ m_V+ c\  m_T}{4
\sqrt{15}}\left(\frac{\Omega}{m_T^2 m_V^2}\right)^2 \right.\nonumber\\
&&+\left.O\left(\left(\frac{\Omega}{m_T^2m_V^2}\right)^3\right)\right]. \eea
\end{widetext}
In the rest frame of the decaying particle $\eta_{c2}$ (denoted by $T$ here), the prefactor
$\sqrt{\Omega}/m_T = |\vec{p}_V|$ is exactly the requirement of the $P$-wave decay of
$\eta_{c2}\rightarrow \gamma J/\psi$. The physical implication of these expressions has been
discussed in context and is omitted here.

\end{document}